\documentclass[twocolumn]{aastex62}

\usepackage{graphicx}
\usepackage{float}

\shorttitle{Mass-Temperature Relation}
\shortauthors{Evans et al.}

\begin{document}


\title{The Mass-Temperature Relation for B and Early A Stars Based on {\it IUE} Spectra of  Detached Eclipsing Binaries}



\correspondingauthor{Nancy Remage Evans}
\email{nevans@cfa.harvard.edu}


\author[0000-0002-4374-075X]{Nancy Remage\ Evans}
\affiliation{Center for Astrophysics $|$ Harvard \& Smithsonian  \\
60 Garden St. \\
Cambridge, MA 02138, USA}

\author[0000-0003-4203-4973]{Mckenzie G.\ Ferrari}
\affiliation{University of Massachusetts-Dartmouth \\
285 Old Westport Rd.  \\
North Dartmouth, MA 02747}

\author[0000-0001-5513-029X]{Joanna Kuraszkiewicz}
\affiliation{Center for Astrophysics $|$ Harvard \& Smithsonian  \\
60 Garden St. \\
Cambridge, MA 02138, USA}

\author{Steven Silverberg}
\affiliation{Center for Astrophysics $|$ Harvard \& Smithsonian  \\
60 Garden St. \\
Cambridge, MA 02138, USA}

\author{Joy Nichols}
\affiliation{Center for Astrophysics $|$ Harvard \& Smithsonian  \\
60 Garden St. \\
Cambridge, MA 02138, USA}

\author{Guillermo Torres}
\affiliation{Center for Astrophysics $|$ Harvard \& Smithsonian  \\
60 Garden St. \\
Cambridge, MA 02138, USA}

\author{Makenzi Fischbach}
\affiliation{Wellesley College  \\
Wellesley, MA }

\begin{abstract}
Ultraviolet spectra were taken of 25 Detached Eclipsing Binaries (DEBs) with spectral 
types O, B, and early A 
with the {\it International Ultraviolet Explorer (IUE)} satellite 
in the 1150 to 1900 \AA\/  region.  The spectra were compared with 
BOSZ model atmospheres (Bohlin, et al. 2017). The composite spectra of the DEBs 
were modeled by a combination of models representing the hot and cool components, and 
the temperatures of the hottest components of the systems were determined.  
 From these temperatures a direct 
Mass-Temperature relation was obtained for stars close to the main sequence with 
solar metallicity  for B and early A stars:

  log $M_{\odot}$ =  -5.90 $\pm$ 0.27 + (1.56 $\pm$  0.07) x log T

 This relation allows a mass to be inferred for comparable stars from
an ultraviolet spectrum.  The five chemically peculiar Am stars in the sample have 
larger radii than normal A stars of the same mass.  


\end{abstract}




\section{Introduction} \label{sec:intro}

Observationally measured masses are critical to understanding stars.  Evolution 
calculations  are tied to these measures.  Eclipsing binary systems where the 
inclination of the system can be determined are a crucial part of the sample of
stars for which we have accurate masses.  Andersen (1991) compiled a list of
detached eclipsing binaries (DEBs) with
accurately determined masses.  This list has subsequently been updated by
Torres, Andersen, and Gimenez (2010;  TAG below).

Well covered photometric and spectroscopic curves analyzed with modern software provide
highly accurate masses, radii, and temperature ratios between the two components 
of the system.  However a valuable parameter in relating these properties to, for 
example, evolutionary tracks is the temperature of each of the two stars.  These must 
be determined from, for instance, spectral types.  

When the Andersen list was drawn up, the {\it International Ultraviolet Explorer} 
(IUE) Satellite was operating.  It provided well calibrated spectra particularly in 
the 1150 to 1900 \AA\/ ultraviolet (UV) region.  For stars of A spectral types and hotter, a large
fraction of the flux is contained in this spectral region.  For this reason a program was 
set up to observe eclipsing systems from the Andersen list with primaries in this
spectral type range.  
Comparison of these spectra with atmospheric models provides
a temperature for the primary.
The goal of this study is to obtain a temperature which 
describes the energy distribution in this important wavelength region.
 This results in a {\it direct} mass-temperature relation 
for hot stars, which  can be used to infer the mass of a star from   
ultraviolet spectra.


\section{Observations} \label{sec:Observ}

The {\it International Ultraviolet Explorer} (IUE) Satellite was launched in 1978 and
was operated by NASA, ESA and the British SERC. 
It provided spectra in the wavelength regions 
 1150 to 1975 \AA\/ and 1910 to 3300 \AA, in two resolution modes (resolution
0.1 and 6 \AA\/ respectively for high and low resolution). 
  Full details of the 
satellite, spectrographs and cameras are given in Harris and Sonneborn (1987).

For eclipsing binaries which have an O, B, or early A component we  
obtained  low resolution spectra with the IUE satellite in the short wavelength region 
through the large aperture.
 These were largely from the program TEPNE (PI: Evans), supplemented  by 
archival data when available.
The observations are summarized in Table~\ref{table:Collection}.  Columns list the variable star 
designation, the HD or BD number, the exposure time,
the spectrum number,   the 
observing date and the observing time, the maximum data count (Data Numbers) 
in the 
spectrum and the background count.  Most spectra were well exposed; however
a maximum count of 255 means some part of the spectrum is saturated and 
cannot be used. 

\begin{deluxetable*}{lllrrrrr}
\tablenum{1}
\tablecaption{\textit{IUE} Satellite Spectra for DEBs \label{table:Collection}}
\tablewidth{900pt}
\tabletypesize{\scriptsize}
\tablehead{
\colhead{System} & \colhead{ID} & \colhead{Exposure (s)} & 
\colhead{Spectrum ID} & \colhead{Obs Date} & 
\colhead{Obs Time} &  \colhead{Max Count} &
\colhead{Background}   \\ 
} 
\startdata
        \textbf{V478 Cyg} & HD 193611 &  427.5  & SWP48207 & 1993/07/22 & 16:50:08  & 72 & 16  \\ 
        \textbf{AH Cep}& HD 216014 & 89.6  & SWP10148 & 1980/09/16 &5:07:28 & 255  & 17 \\
       \textbf{V578 Mon} & HD 259135 & 2699.5 &  SWP17905 & 1982/09/10 & 9:21:56  & 178 & 58  \\ 
       \textbf{QX Car} &  HD 86118  & 4.8 &  SWP50174 & 1994/03/06 & 22:20:14 &  179 & 14  \\ 
       \textbf{V539 Ara}& HD 161783 & 2.7  &  SWP48476 & 1993/08/27 & 16:25:07 &  132 & 16  \\
        \textbf{CV Vel}& HD 77464 &  4.4  & SWP50160 & 1994/03/03 & 20:46:04 &  118 & 16  \\
            \textbf{U Oph }& HD 156247 &  10.5 &  SWP48498 &  1993/08/29 & 15:14:27 &  189 & 15  \\
        \textbf{V760 Sco}& HD 147683 &  41.6   & SWP48497 & 1993/08/29 & 14:10:13 &  138 & 15  \\
        \textbf{GG Lup}& HD 135876 &  8.5 & SWP48496 & 1993/08/29 & 12:58:18 &  255 & 17  \\
         \textbf{$\zeta$ Phe}& HD 6882 &  1.1 & SWP48495 & 1993/08/29 & 11:37:37  & 143  & 16  \\
        \textbf{$\chi^{2}$ Hya}& HD 96314 &  6.8  & SWP50685 &  1994/05/03 & 20:28:43  & 108 & 15  \\ 
        \textbf{IQ Per} & HD 24909 &  111.7  & SWP50177 & 1994/03/07 & 2:19:11   & 192 & 16  \\
        \textbf{PV Cas}& HD 240208 &  880.5 &  SWP48206 & 1993/07/22 & 15:43:50  & 99 & 19  \\
        \textbf{V451 Oph}& HD 170470 &  208.8  &  SWP48217 &  1993/07/23 & 18:52:14   & 181 & 16  \\
        \textbf{WX Cep }& HD 213631 &  2999.1   & SWP50179 & 1994/03/07 & 20:38:38   & 109 & 19  \\ 
        \textbf{TZ Men}& HD 39780 & 29.8 &  SWP50175 & 1994/03/06 & 23:26:55 &  164 & 15  \\
        \textbf{V1031 Ori}& HD 38735 &  69.5  & SWP50176 & 1994/03/07 & 0:39:52   & 132 & 15  \\ 
       \textbf{$\beta$ Aur}& HD 40183 & 0.7  & SWP50172 &  1994/03/06 & 19:34:58   & 114 & 14 \\
       \textbf{YZ Cas}& HD 4161 & 39.6  & SWP26934 &  1985/10/14 & 14:00:51  & 183 & 18 \\
        \textbf{V624 Her}& HD 161321 &  85.5  & SWP48216 &  1993/07/23 & 17:35:04   & 104 & 16  \\ 
        \textbf{GZ CMa} & HD 56429 &  419.7  & SWP50173 &  1994/03/06 & 20:55:47   & 126 & 16  \\ 
        \textbf{V1647 Sgr} & HD 163708 &  99.8  & SWP48475 & 1993/08/27 & 15:35:53  & 150 & 15  \\ 
        \textbf{EE Peg}& HD 206155 &  119.5  & SWP50684 & 1994/05/03 & 19:04:41  & 139 & 17  \\
        \textbf{VV Pyx}& HD 71581 &  77.7  & SWP50161 & 1994/03/03 & 22:08:02  & 189 & 15  \\ 
        \textbf{AY Cam}& BD+77$^{\circ}$ 328 & 2699.5  & SWP29742 &  1986/11/24 & 1:41:45 &  63 & 21  \\
\enddata
\end{deluxetable*}

Table~\ref{table:phas} provides the orbital phases at which the {\it IUE} observations
were obtained.  Columns list the system, the period, the HJD of primary minimum,
the reference for the orbit, the phase of secondary minimum, the HJD of the {\it IUE}
observation, the orbital phase of the {\it IUE} observation, and notes.  Most of
the  {\it IUE} observation were taken outside of eclipse.  In the final column
asterisks indicate 4 systems for which the observation occurred 
 during some part of an eclipse.  Three
of these systems
(V478 Cyg, AH Cep, and V587 Mon) were discarded during spectral fitting, as 
discussed in Section~\ref{systems}.  V539 Ara was also mildly affected by an eclipse,
as is also discussed in Section~\ref{systems}.

\begin{deluxetable*}{lllrrrrr}
\tablenum{2}
\tablecaption{\textit{Phases of Observation}  \label{table:phas}}
\tablewidth{900pt}
\tabletypesize{\scriptsize}
\tablehead{
\colhead{System} & \colhead{P} & \colhead{Min I} & 
\colhead{Ref} & \colhead{Phase of Min II} & 
\colhead{{\it IUE} } &  \colhead{{\it IUE} Phase} &
\colhead{Notes}   \\ 
\colhead{} & \colhead{(d)} & \colhead{HJD} & \colhead{} & 
\colhead{} & \colhead{HJD} & \colhead{} &
\colhead{}  \\
} 
\startdata
V478 Cyg  & 2.88090063  & 2444777.4852  &   1    &   0.5067  &      2449191.2014 &  0.0613 & * \\ 
AH Cep   &  1.774761   &  2445962.7359  &   2   &    0.5000	&       2444498.7135 &  0.0876 & * \\ 
V578 Mon  & 2.40848    &  2449360.625   &   3    &   0.4507	 &      2445222.8902 &  0.0141 & * \\ 
QX Car    & 4.4780399  &  2443343.41512  &  4    &   0.4010	  &     2449418.4307 &  0.6238 &  \\
V539 Ara  & 3.169112   &  2445056.777    &  5    &   0.3307	  &     2449227.1841 &  0.9545 &  *  \\  
CV Vel   &  6.889494   &  2442048.66894  &  6     &  0.5000	  &     2449415.3653 &  0.2652  &  \\
U Oph    &  1.6773458  &  2442621.6212   &  7    &   0.5017	  &     2449229.1350 &  0.2675  &  \\
V760 Sco &  1.7309295  &  2443250.8268   &  8    &   0.5101	  &     2449229.0904 &  0.7881  &  \\
GG Lup    & 1.84960    &  2446136.7398   &  9    &   0.5064	  &     2449229.0404 &  0.8754  &  \\
$\zeta$ Phe  & 1.6697724 &    2441643.6890 &   10  &     0.5000	   &    2449228.9844 &  0.7122  &  \\
$\chi^2$Hya & 2.267701  &   2442848.6107  &  11   &    0.5000	   &    2449476.3532 &  0.6704  &  \\
IQ Per    & 1.74356214 &  2444290.3664  &  12    &   0.5222	   &    2449418.5966 &  0.2374   &  \\
PV Cas    & 1.75047    &  2449210.7956  &  13    &   0.5027	   &    2449191.1554 &  0.7801  &  \\
V451 Oph  & 2.19659557 &  2445886.53335 &  14    &   0.4978	   &    2449192.2862 &  0.9438  &  \\
WX Cep   &  3.3784535  &  2425088.537   &  15    &   0.5000	   &    2449419.3601 &  0.7635  &  \\
TZ Men   &  8.56900    &  2442403.7085  &  16    &   0.5096	   &    2449418.4770 &  0.6216  &  \\
V1031 Ori & 3.405565   &  2444643.665   &  17    &   0.5000	   &    2449418.5276 &  0.0765  &  \\
$\beta$ Aur  & 3.96004673  & 2452827.195693  & 18  &     0.5000	    &   2449418.3159 &  0.1819  &  \\
YZ Cas    & 4.46722236 &  2445583.78664 &  19    &   0.5000	   &    2446353.0839 &  0.2093  &  \\
V624 Her  & 3.894977   &  2440321.005   &  20    &   0.5000	   &    2449192.2326 &  0.6072  &  \\
GZ CMa   &  4.8008500  &  2443581.56132  & 21    &   0.5000	   &    2449418.3720 &  0.7869  &  \\
V1647 Sgr &  3.28279251 &  2441829.69510 &  22  &     0.2621	   &    2449227.1499 &  0.4031  &  \\
EE Peg   &  2.62821423 &  2440286.4349  &  23    &   0.5000	   &    2449476.2949 &  0.6176  &  \\
VV Pyx   &  4.5961832  &  2444620.65895 &  24    &   0.4804	   &    2449415.4222 &  0.2054  &  \\
AY Cam   &  2.73496794  & 2443572.74441 &  25    &   0.5000	   &    2446758.5706 &  0.8496  &  \\
\enddata
\tablecomments{*Starred systems are where  {\it IUE} observations overlap eclipse;
References: 1. Wolf et al. 2006; 2. Holmgren et al. 1990; 3. Hensberge et al. 2000;
4. Andersen et al. 1983; 5. Clausen 1996; 6. Yakut et al. 2007; 7. Vaz et al. 2007;
8. Andersen et al. 1985; 9. Andersen et al. 1993; 10.  Andersen et al. 1983;  11. Clausen and
Nordstrom 1978; 12.  Wolf et al. 2006; 13. Barembaum \& Etzel 1995; 14. Clausen et al. 1986;
15. Popper 1987; 16.  Andersen et al. 1987; 17. Andersen et al. 1990; 18. Southworth et al.
2007; 19. Pavlovski et al. 2014; 20. Popper 1984; 21. Popper et al. 1985; 22. Andersen \& Gimenez
1985; 23. Lacy \& Popper 1984; 24. Andersen et al. 1984; 25. Williamon et al. 2004}


\end{deluxetable*}


\section{Data Reduction}

Spectra were retrieved from the Hubble Space Telescope (HST) Mikulski  
Archive for Space Telescope  (MAST)
archive, having been processed with the NEWSIPS pipeline (Nichols and Linsky 1996).
Further processing was done with the IUE Regional Analysis Facility (IUERDAF) 
software package installed at the Harvard \& Smithsonian Center for 
Astrophysics, High Energy Division.     
The correction to the IUE fluxes derived by Bohlin and Bianchi (2017) to match the HST
CALSPEC files (e.g. Space Telescope Imaging Spectrograph [STIS] spectra) 
was applied to the IUE fluxes.


IUE NEWSIPS  pipeline already removes known features such as the 
reseaux marks on the camera and 
cosmic ray hits.  In addition, for most 
exposure times geocoronal Lyman $\alpha$ emission is present in the line cores, which was 
removed (``blemished'') by eye.
The spectra were then dereddened using the reddening law of  Cardelli, Clayton, and 
Mathis (1989) using the  IDL
 procedure unred\_ccm.pro.  E(B-V) was 
taken from TAG (except for WX Cep)
 and is listed in the summary table (Table~\ref{table:Binaries}).  The reddening 
for WX Cep was taken from 
Graczyk, et al. (2019).  

\subsection{BOSZ Model Atmospheres} \label{sec:BOSZ}
A grid of stellar atmospheres has been computed by Bohlin et al. (2017)
 from the Kurucz Atlas9 code, which are available in the MAST archive.




To cover a range of hot stars, we used an initial 58 models spanning a temperature range
 of 5,000--35,000 K. For the BOSZ models used in this work, we chose the following
 initial conditions appropriate to main sequence stars in \textit{IUE} low resolution spectra:
 surface gravity (log g) = 4.0, solar metallicity [M/H] = 0,  solar carbon abundance [C/M] = 
 solar, alpha abundance [alpha/M] = 0, instrumental broadening = 500 km/s, and 
microturbulence = 2 km/s. 
 The selection of these parameters was to reflect the characteristics of the DEBs 
and {\it IUE} spectra as 
closely as possible, while keeping the analysis uniform for  all systems.

While the BOSZ temperature models span a large temperature range,
 the increments at which a temperature model is available varies throughout the range. 
For cooler temperatures, models are available every 250~K. For warmer temperatures, 
models are available every 500 K. At the hottest temperatures, models are available 
every 1000 K.
To determine more precise temperature uncertainties, we generated 
linearly interpolated   models
 from the original BOSZ models. 


\section{Analysis} \label{sec:Analysis}

The goal of this study was to determine a temperature which best reproduced the spectrum from 1150
to 1975 \AA\/ within the framework of the models. 
The most important aspects of the analysis approach are:

$\bullet$ The composite spectrum of the binary  was modeled with  a combination of spectra from the 
models of the hot and cool components. 

$\bullet$ The hotter star in the binary dominates at the shortest wavelengths, and only that 
temperature is determined from the fits.

$\bullet$ The analysis is based on best-determined parameters from the eclipse solution: 
the mass M, 
the radius R, and the ratio of temperatures of the two stars T$_1$/T$_2$ where 1 and 2 refer
to the primary  (most massive) and secondary of the binary system. 

 The details of the matching process are described in the sections below.

\subsection{Spectrum--Model Comparisons} \label{sec:Compar}

The comparison between the IUE spectrum and the BOSZ models was made as follows.  The radii of 
the primary and secondary stars in the systems are well determined from the eclipse solutions.  The 
analysis was started with the temperatures T$_1$ and T$_2$ from spectral types from TAG.  
A model atmosphere  was selected close to
the temperature for each star, and then the model of the secondary was scaled using 
$(R_2 /R_1)^2$ from the eclipse solutions. These two models were then summed and normalized 
to the flux of the IUE spectrum.  For the hottest stars, the Ly $\alpha$ region
  (1180 to 1250~\AA) was excluded because of contamination by interstellar Ly $\alpha$
  absorption.
The summed model flux and the IUE flux were then used to calculate
a standard deviation (SD). The hottest star (generally the primary) dominates heavily at the 
shortest wavelengths, so a series of comparisons were made stepping  the 
temperature of the primary  T$_1$ through a series of temperatures bracketing the temperature 
expected from TAG.  These comparisons were used in two ways.  First, the comparison was 
inspected as in Fig.~\ref{v451spect}.  In general, this inspection showed differences between models
of  T$_1$ of  100 to 200 K for the cooler primaries and 500 K for hotter primaries.  
Second the difference  between the composite model and  the {\it IUE} spectrum 
was formed and plotted through the sequence of T$_1$ 
(Fig.~\ref{v451diff}).
  Again, in general, differences were clear between temperatures (T$_1$) of about 200 K, particularly
at the long and short wavelength regions of the spectra.  The standard deviations produced
a parabola as a function of T$_1$ (Fig.~\ref{v451par}), 
from which the minimum was used to determine  T$_1$.

  Visual inspection of the spectral 
  comparison and the difference plots was one approach to
  temperature determination.
In this case, an uncertainty was estimated from the 
temperatures which could be identified by eye as too high and
too low to be matches ($\delta$ T$_1$).  
 The temperature was assumed to be between these temperatures, or a quarter of 
$\delta$ T$_1$.





We also tried a $\chi^2$ approach, using the {\it IUE} instrumental error. 
However, because the study covered a large range of temperatures from 30,000 to 7,000 K two effects
complicated the  $\chi^2$ estimation. 
First, different wavelength regions were very differently exposed on the camera, and hence 
had very different distributions of actual uncertainties.  Second the line opacity  
is very different over that temperature range. Both these were difficult to incorporate into the
analysis.   Standard deviations seemed to give 
more consistent temperature  uncertainties throughout the range.

Spectral comparisons such as Fig.~\ref{v451spect} show the full spectrum.  However, portions of 
the spectrum which are saturated are marked in the upper left spectrum, as are reseau marks, 
and geocoronal Ly $\alpha$.  While these regions are shown in the figure, they are omitted 
from the SD calculation.  In addition in cool spectra, there are features in the models near 
1550\AA, and  1700\AA.  While these appear in the models which have  complex opacity, they are 
not seen in any of the spectra, so they are omitted from the fitting (Fig.~\ref{yzcasspect}).


\subsection{Composite Spectra: Groups}
The detailed analysis was done in four groups.

 \begin{itemize}

\item Group I:  The primary and secondary have fairly similar temperatures

\item Group II: The secondary is markedly cooler than the primary

\item Group III: The secondary is actually hotter than the primary, i.e. the 
primary has evolved off the main sequence   

\item Group IV: What is it?

\end{itemize}

{\bf Group II}: These are the easiest to analyze.  Because the secondary is significantly cooler than the 
primary, once the model flux is scaled $(R_2 /R_1)^2$, the contribution of the secondary is very 
small.  This means that a modest change in temperature T$_2$ will have an insignificant effect
 on the summed model spectrum.  Hence the procedure stepping through T$_1$ values will produce
an optimal value of T$_1$.

{\bf Group I}: The analysis is a little more complicated.  It begins as for Group II to determine
a  T$_1$ using an assumed  T$_2$.   (Solutions for which  
 T$_2$ was larger than  T$_1$ were discarded.) 
An additional constraint was then imposed.  Since the 
temperature ratio T$_1$/T$_2$ is well determined from the eclipse solution, it can be 
used to obtain T$_2$ from 
 the value of T$_1$ determined in the first step.  With this new value of 
T$_2$ the process is repeated until a solution converges near the T$_1$/T$_2$ 
from the eclipse solution.   Occasionally, the 
 temperatures of the two components are determined to be indistinguishable.

{\bf Group III}: This group similarly requires iteration, although in this case the ratio of the radii brings
the models of the two temperatures closer together, complicating the determination of the temperature
of the hotter star.

{\bf Group IV}: This group contains puzzling or otherwise poorly defined solutions.

\subsection{Examples}

The analysis process is illustrated with examples from a hot star in the sample V451 Oph, and a 
cool star YZ Cas, both in  Group II.

\begin{figure*}
\plotone{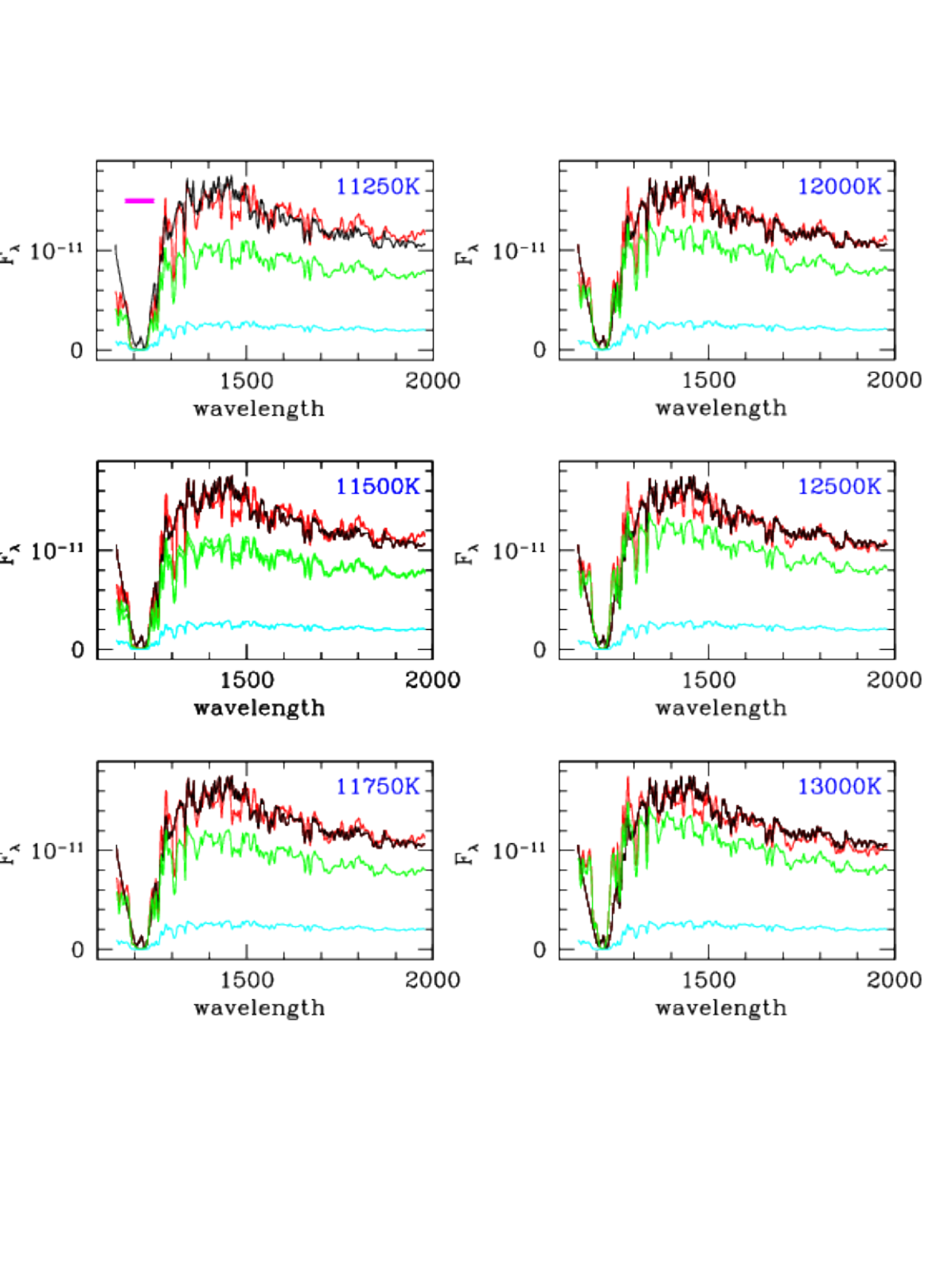}
  \caption{V451 Oph: Comparison of the {\it IUE}  unreddened
  spectrum and models.  Lines are: {\it IUE} 
spectrum: black; hot model: green; cool spectrum: blue;  
 model composite: red.  Model T$_1$ (hot) temperatures are indicated in 
each panel. The best fit is for T$_1$ = 12000~K.
Wavelength is in \AA; flux is in ergs~cm$^{-2}$~s$^{-1}$~\AA$^{-1}$ in all
figures.
\label{v451spect}}
\end{figure*}

\begin{figure}
\plotone{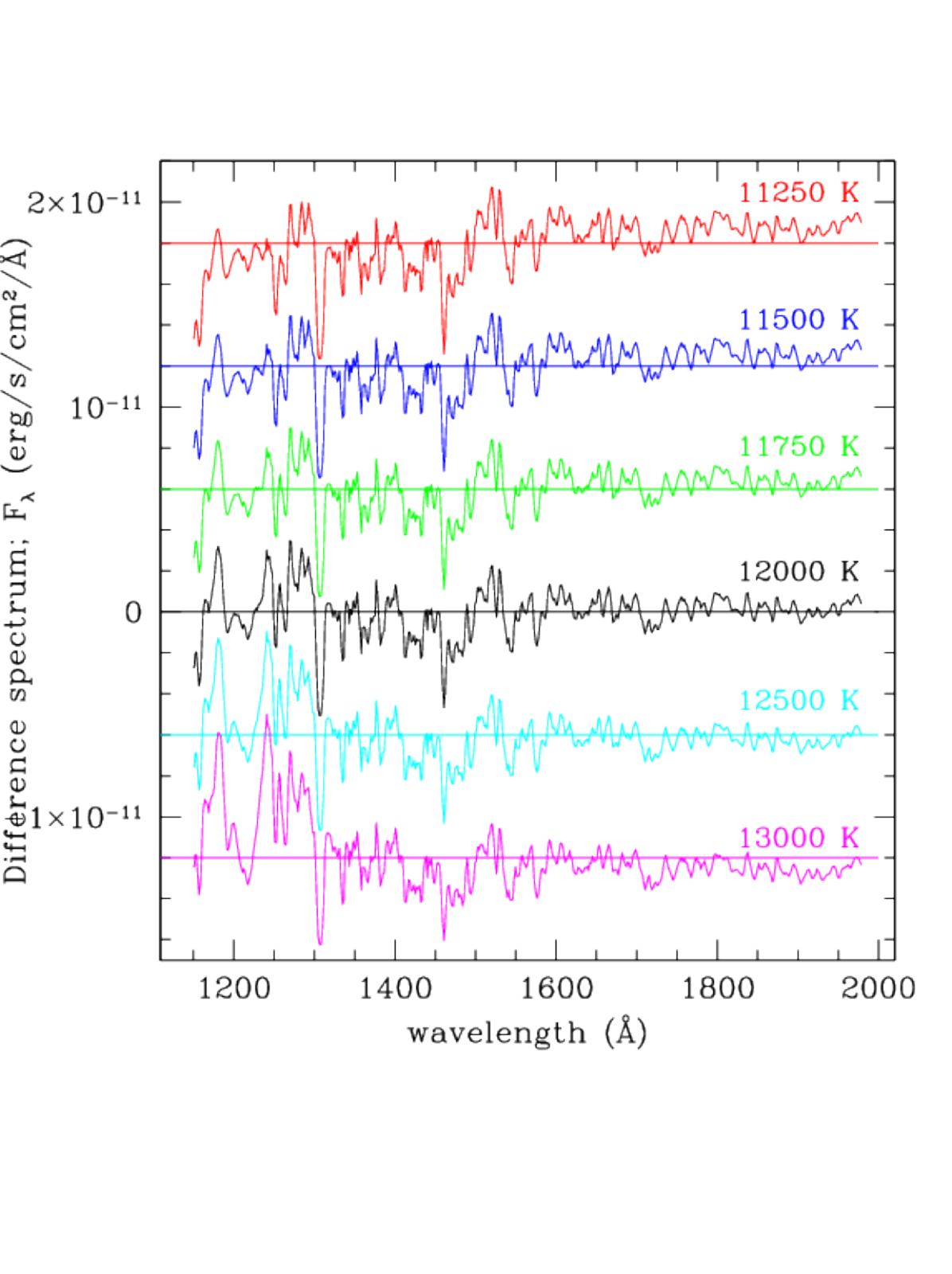}
\caption{V451 Oph: The difference between the composite model and  
the spectrum.  
The differences from models of six T$_1$ temperatures (indicated in
the figure) are shown from top to 
bottom.  An arbitrary constant has been added to the difference spectra so
  that they do not overlap.
(T$_2$ fixed at 9800 K.)  
The best fit is indicated 
with a black line.  The changes as a function of model temperature 
are clear in the comparison.   The coolest model (top)  is too 
low in flux for   the shorter wavelengths as compared with the (normalized) longer
wavelengths. Conversely, for a hotter model (bottom)  the model fluxes 
are larger than the spectrum for wavelengths shorter than 1300 \AA.
  In this temperature range, the poor agreement in the 
Ly $\alpha$ region near 1200 \AA\/ is particularly marked for the
hottest model.   
\label{v451diff}}
\end{figure}

\begin{figure}
\plotone{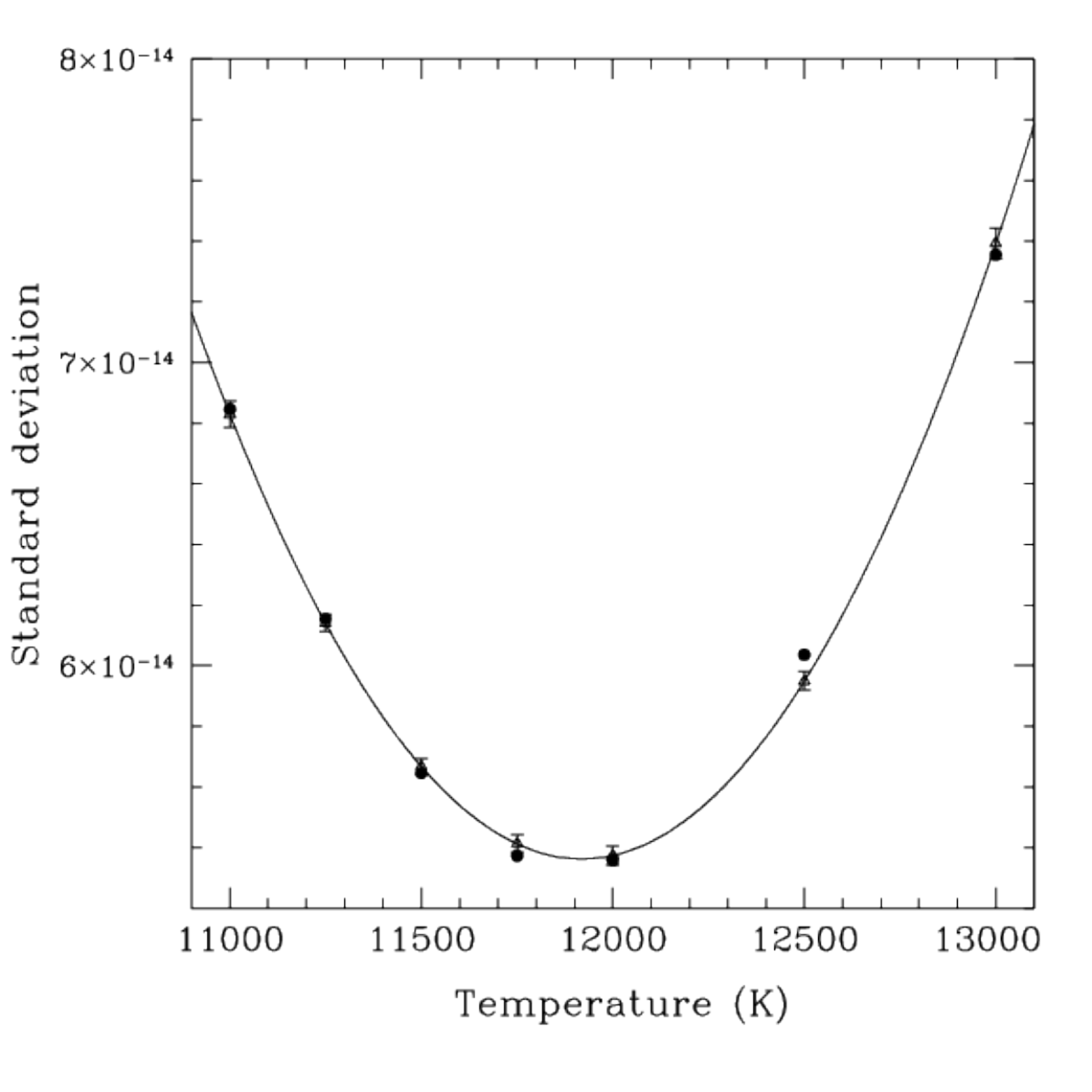}
\caption{V451 Oph: The temperature determination: the standard deviation
as a function of temperature of the primary.  The temperature of the 
secondary was fixed at 11000~K; dots: the standard deviation from the
spectrum-model comparison; triangles: the parabola fit.
\label{v451par}}
\end{figure}

{\bf V451 Oph}   
The spectrum/model comparison, the difference comparison, and standard deviation parabola are
shown in Figs.~\ref{v451spect},  ~\ref{v451diff}, and  ~\ref{v451par} respectively.
In Fig~\ref{v451diff} since the temperatures for the primary 11500 and 12500~K are 
clearly ruled out, an estimate of the ``visual'' uncertainty for T$_ 1$ is 250~K.

\begin{figure*}
\plotone{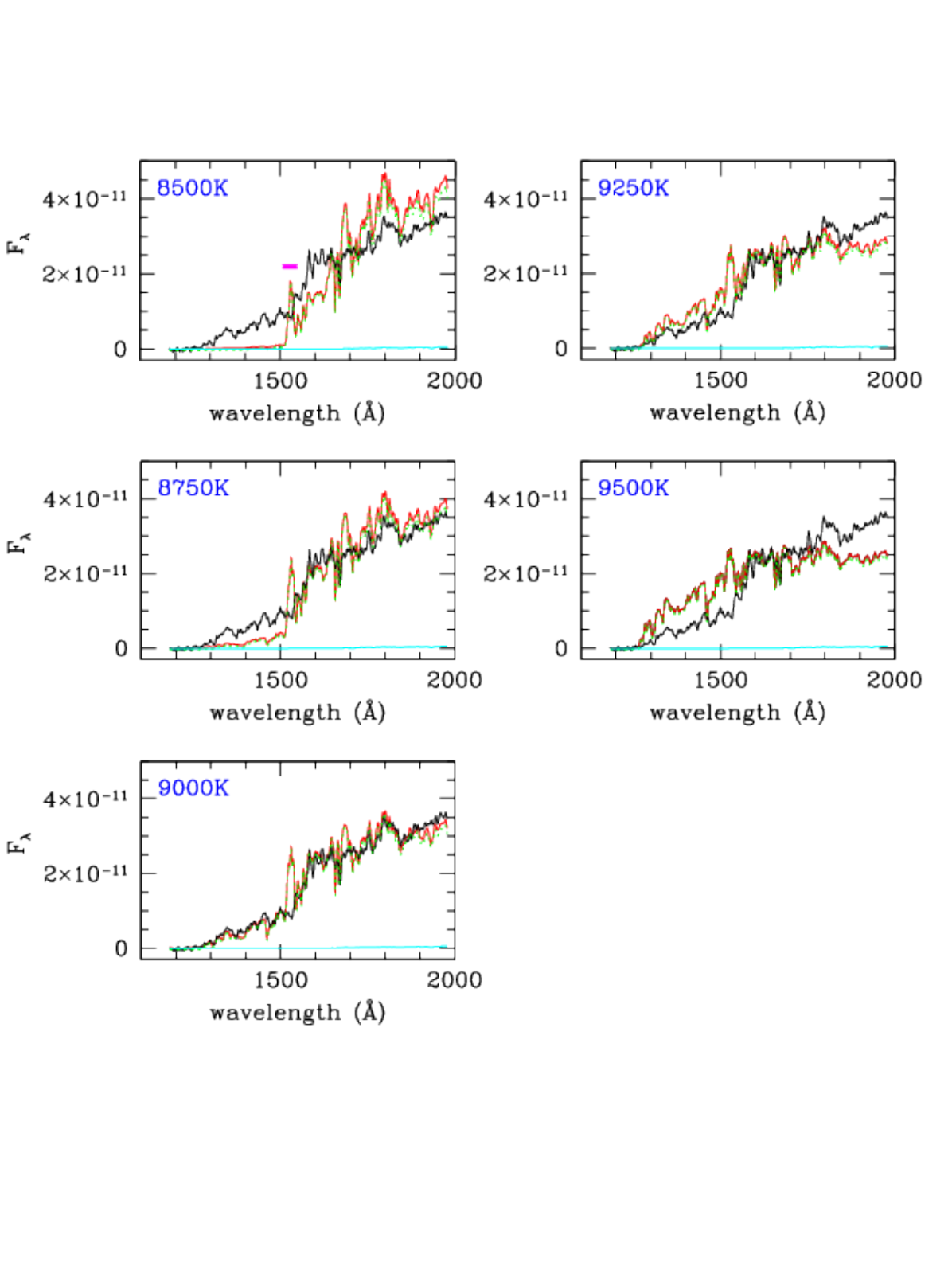}
\caption{YZ Cas: Comparison of {\it IUE} spectrum and models.  Lines are the same 
as in Fig~\ref{v451spect}.  The magenta line above the spectrum in the
upper left at 1550 \AA\/ indicates that this feature was not 
included in the fit.  Best fit is for T$_1$ = 9000~K.
\label{yzcasspect}}
\end{figure*}

\begin{figure}
\plotone{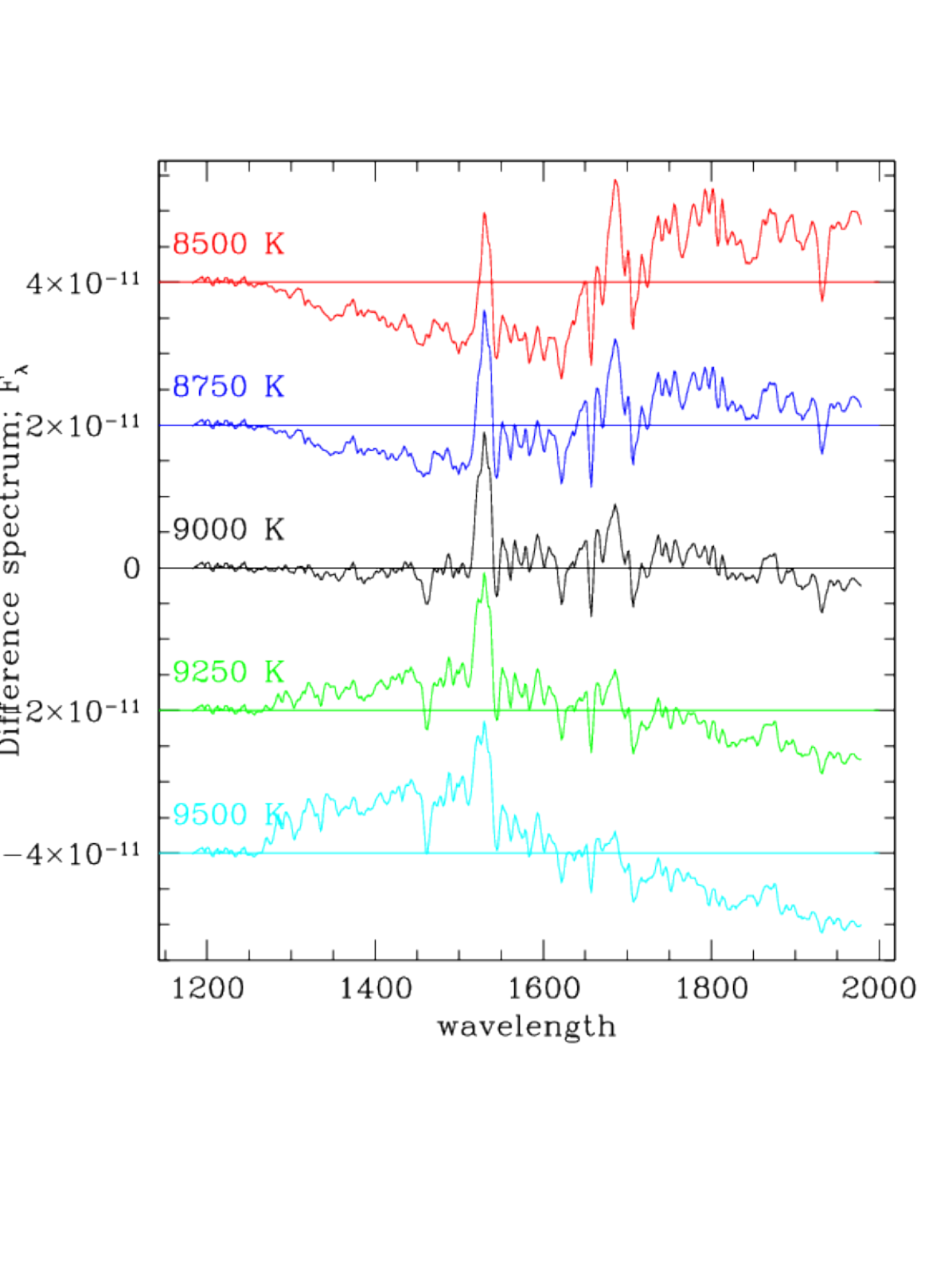}
\caption{YZ Cas: The difference between the model and 
the {\it IUE} spectrum.  
The differences from models of five temperatures is shown from top to 
bottom with increasing model temperature.  The best fit is indicated 
with a black line.     
\label{yzcasdiff}}
\end{figure}

\begin{figure}
\plotone{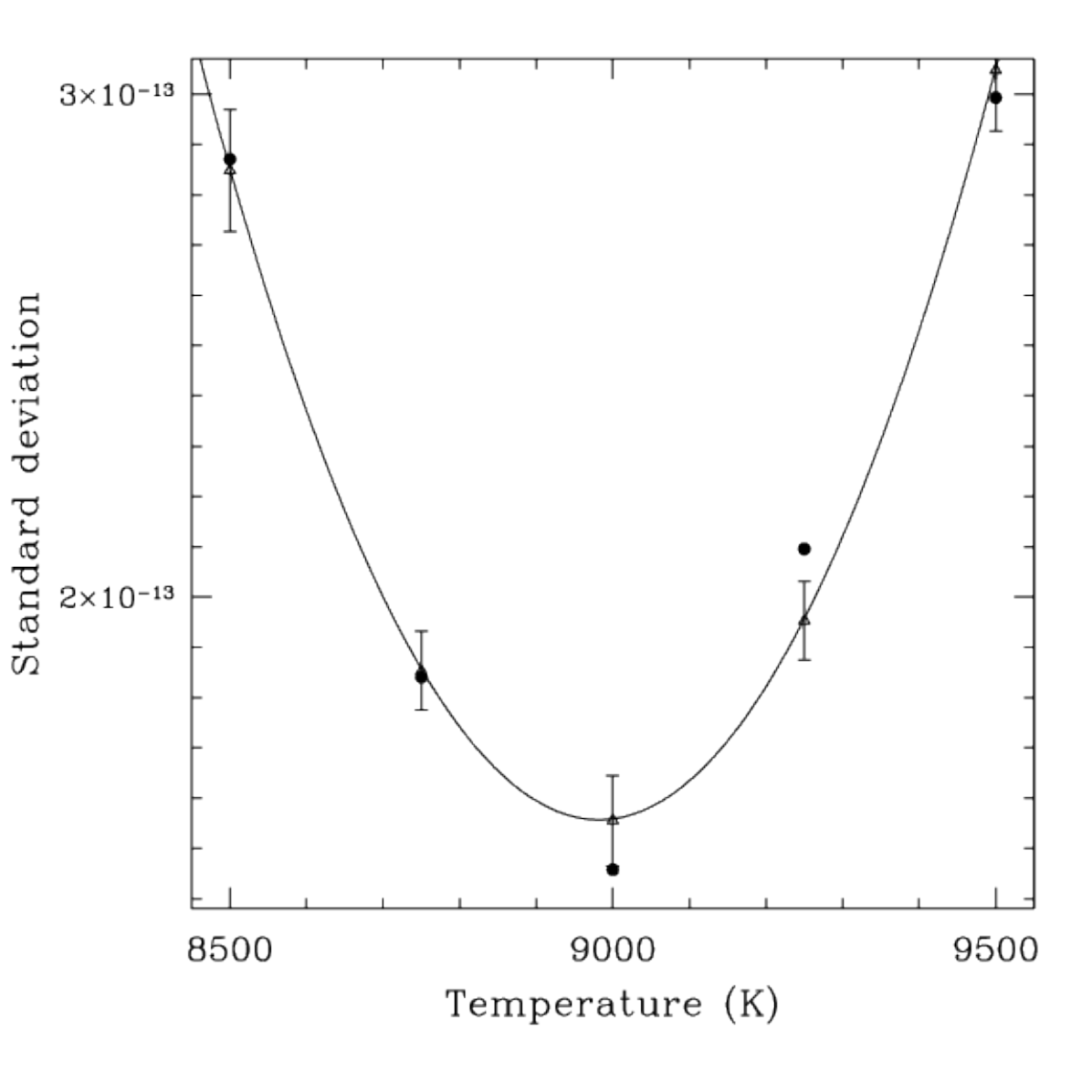}
\caption{YZ Cas: The temperature determination: the standard deviation
as a function of temperature of the primary.  The temperature of the 
secondary was fixed at 7250 K; dots: the standard deviation from the
spectrum-model fit; triangles: the parabola fit
\label{yzcaspar}}
\end{figure}

{\bf YZ Cas}   
As an example of a cooler star YZ Cas is used.  The spectrum/model comparison, 
the difference comparison, and the standard deviation parabola are 
shown in Figs.~\ref{yzcasspect}, \ref{yzcasdiff} and   \ref{yzcaspar}. 
From Fig.~\ref{yzcasdiff}, the temperatures 8750 and 9250 K are 
ruled out, making an estimate on the visual uncertainty of 125 K for  T$_1$. 
The temperature determination from the SD parabola is shown in Fig.~\ref{yzcaspar}


\subsection{Sources With One-Temperature Fits}

For PV Cas, the temperatures, masses, and radii of both the primary and secondary are assumed 
to be the same (TAG). In this case, only one temperature T$_1$ was fit to the model.
 A 1T fit was also explored in other cases, specifically for CV Vel, QX Car,
  and VV Pyx.  



\subsection{Reddening Tests}
One of the first concerns in working with stellar energy distributions (SEDs) in the 
ultraviolet is the effect of reddening on derived temperatures.  While the ultraviolet 
fluxes are heavily affected by reddening or uncertainty in reddening, temperatures 
derived from ultraviolet SEDs are surprisingly robust to changes in reddening, at least 
for the temperatures of the stars that we are working with.  This is for two reasons. 
First the wavelength range we are working with is only 750 \AA~
(1150 to 1900 \AA), making the variation 
of flux with a change in E(B-V) {\it within} that wavelength range
quite small.  Second, the observed energy 
distributions have a well defined shape in flux versus wavelength so that the comparison with 
the models is fairly tightly defined.   For example,  the hottest stars typically
have  flux  which
decreases sharply at shorter wavelengths.

We have tested the sensitivity of our temperature fits 
to reddening on a  range of E(B-V) values for  V451 Oph
(Table~\ref{ebv}).  The uncertainties in Table~\ref{ebv} are from 
the visual estimates as discussed in Sect. 4.1.
Although the range in E(B-V)
is much larger than possible uncertainties in E(B-V),  the resulting 
temperatures only differ by  $\pm$ 450 K or  $\pm$ 4\%.  Thus for a typical 
uncertainty in E(B-V) (perhaps 0.03 mag) the uncertainty in the temperature is 
$<$1\%.
In addition, most values of  E(B-V) for the targets are
very small (Table~\ref{table:Binaries}),
implying small uncertainties in the reddening and temperature. 

\begin{deluxetable}{rrr}
\tablenum{3}
\tablecaption{E(B-V) Tests for V451 Oph \label{ebv}}
\tablewidth{900pt}
\tablehead{
\colhead{E(B-V)} & \colhead{T} & \colhead{$\pm$} \\ 
\colhead{(mag)} & \colhead{(K)} & \colhead{(K)}  \\ 
} 
\startdata
  0.05 &  11734.9 & 125  \\
   0.10 &  11933.4 & 125   \\
   0.158  & 12161.8   & 250  \\
   0.20 &  12376.3 &  312  \\
   0.25 &  12639.4 &  280   \\
\enddata
\end{deluxetable}

\subsection{M-K Standards}
In order to explore the sensitivity of the fits to IUE spectra of early B stars, three 
spectra from the IUE Ultraviolet Spectral Atlas (Wu, et al. 1983) of Morgan-Keenan 
spectral type standards
(or similar substitutes where necessary) were analyzed in the same way as the DEBs. 
Results are provided in Appendix A.   For comparison temperatures for several 
calibrations of the MK classes are provided.  

\subsection{Temperature Summary}

The temperature results are summarized by Group in Table~\ref{t.group}.
The columns list the star and T$_1$ from the parabola fit. The two error 
estimates are from the standard deviation parabola and the visual 
inspection of the spectrum/model comparison.  T$_2$ is the temperature 
of the secondary in the fit, starting from the value in TAG, but 
adjusted if required by the comparisons.  The T$_2$/T$_1$ ratio 
from TAG is in the 
next column.  In borderline cases there is a little overlap in this
ratio between Group I and Group II.  The final column provides notes
to our fits. For
PV Cas a one-temperature fit was made (Sect. 4.4).
For Group III (bottom) T$_2$ is the hotter temperature and is listed 
in Column 2 rather than Column 5.  

 One of the goals of this study was to determine the temperature range
  for which IUE spectra provide accurate temperatures. This is discussed further in
Section~\ref{hot}.


\begin{deluxetable*}{lllrrrr}
\tablenum{4}
\tablecaption{Temperatures by Group \label{t.group}}
\tablewidth{900pt}
\tablehead{
\colhead{Star} & \colhead{T$_1$} & \colhead{Err} & \colhead{Err} &
\colhead{T$_2$} & \colhead{T$_2$/T$_1$ $^a$} & \colhead{Notes $^b$ } \\ 
\colhead{} & \colhead{ } & 
\colhead{ SD} & \colhead{ Vis } & \colhead{ } & & \\ 
\colhead{} & \colhead{(K)} & 
\colhead{(K)} & \colhead{(K) } & \colhead{ (K) }  & &  \\ 
} 
\startdata
{\bf Group II}  & & & & & & \\
 & & & & & & \\
V539 Ara & 19248   &   363 &  500  & 17500 &   0.945   &  \\ 
GG Lup &   15793  &   1518   &  250  &   11500 & 0.746   &  \\ 
$\zeta$ Phe &  14483  &  2440   & 250  & 12000  & 0.833   &  \\ 
IQ Per &   13241 &    3051 &  250  &  8750  &  0.626  & \\
TZ Men  &  10418 &   1123 &  100 & 7250  &  0.692 & \\
V451 Oph &  11980 &    130 & 250  &  11000 &   0.907   & \\
YZ Cas &   8982 &     385 & 125 &  7250   &    0.706  &  \\ 
V624 Her &  8160  &     335     &   125     &  8000  &  0.975 & \\  
GZ CMa &   8566  &   440   &  125    & 8250  &    0.966     & \\ 
EE Peg &   8687  &   119   & 125  & 6500   &  0.741 & \\ 
  & & & & & & \\
{\bf Group I}  &  & & & & & \\
  & & & & & & \\
QX Car &  23428  & --- &  1000  & 23000 &   0.950 & \\ 
CV Vel &  18045  &   563 & 500 & 18000 & 0.989  & \\ 
V760 Sco &  17092 &   854 & 500   &  16000  &  0.964 & \\
PV Cas &   10632  &  677  &  250    &       &  0.999    &  1T  \\
$\beta$ Aur & 9177   &   914 & 125 &  9000 & 0.984  & \\ 
V1647 Sgr &  9361  &    535 &  125   & 8750 &  0.948  & \\
VV Pyx &   9751  &    782 &  125   &  9500 & 1.00  & \\ 
  & & & & & & \\
{\bf Group IV}   & & & & & & \\
 $\chi^2$ Hya & 12160 & 1370 & 250 & 11500 & 0.945 &  \\
  & & & & & & \\
{\bf Group III }  & & & & & & \\
  & & & & & & \\
  &             T$_2$   &  Err   &  Err &     T$_1$  &  & \\   
  &                   &  SD  &   Vis          & &  & \\ 
WX Cep     &    9087 & 1092  &   125  &   8375 &   1.08 & \\
V1031 Ori & 8906  &  419  &  100  & 7750  &  1.07 & \\    
AY Cam  &  7590  &   318  & 150 &  7200 & 1.02 &   \\  
  & & & & & & \\
\enddata
\tablecomments{$^a$ The T$_2$/T$_1$ is from TAG; 
$^b$  1T: one temperature model fit}
\end{deluxetable*}




\section{Comments on Individual Systems} \label{systems}

\subsection{V478 Cyg}  \label{V478_Cyg}
The temperature for V478 Cyg has one of the largest uncertainties in this study
for two reasons.  First, it is one of the weakest exposures.  In addition to 
that the E(B-V) is somewhat uncertain.  The E(B-V) in TAG is 0.85, 
taken from Popper and Etzel (1981).  As they discuss, the B-V 
measurement of the system 
(Popper and Dumont 1977) was done with two telescopes (Palomar and Kitt Peak).  
The larger diaphragm at Palomar included an additional star.  This star is 
4.26 mag fainter than V478 Cyg, hence would have only a very small effect on
the color, most likely making it slightly redder.   Since photometry of the 
system is typically made differentially, measurements of B-V are scarce. 
We were not able to achieve a definitive  fit, and omit V478 Cyg from 
further consideration.

\subsection{AH Cep}  \label{AH_Cep}
 The {\it IUE} spectrum of AH Cep is saturated for half the wavelength region. 
For this reason, the standard temperature fitting is unreliable.    AH Cep is omitted from Mass-Temperature 
fitting.   




\subsection{U Oph}  \label{U_Oph}   Although the exposure of U Oph is reasonably
good, the spectrum is saturated in several important wavelength regions. For
this reason, it is omitted from temperature discussions.

 \subsection{QX Car}  \label{QX_Car}  
  The temperatures of the two components are similar (Group I), but the
  masses are clearly not identical (TAG).  We have investigated both
  1T and 2T fits.  The temperature from the 2T fit is 23430 K, with an
  estimated visual uncertainty of 1000 K.

 \subsection{V578 Mon}  \label{V578_Mon} 
             Because the spectrum rises monotonically toward short wavelengths,
the temperature and reddening are degenerate,  therefore         the temperature
 from the {\it IUE} spectrum is not definitive, and is not used in subsequent 
temperature discussions.   
Subsequent to TAG, a small adjustment was made to the ratio of radii (Garcia, et al. 
2013), resulting in a difference of the normalization factor (R$_2$/ R$_1$)$^2$ of 2\%,
which has no effect on T$_1$.

  \subsection{V539 Ara}  \label{V539_Ara}
  The {\it IUE} observation occurred just inside the start of primary eclipse.
  However, V539 Ara is a Group II system, with the secondary significantly less
  luminous than the primary in the ultraviolet.  The decrease in flux in the
  primary of approximately 5\% will have minimal effect
  on the spectral fitting.  (Fig. ~\ref{v451spect}
  illustrates a similar system.)




\subsection{Group III:  WX Cep,  and V1031 Ori, and  AY Cam} For the 
three stars in this group  the hotter 
star is actually the secondary which is less luminous, and smaller, so    
 the more massive star is less dominant in the ultraviolet.  For this reason
it is more difficult to disentangle the composite 
spectrum, and the temperature of the hotter star is less well determined.  We include 
estimates of the temperatures in Tables~\ref{t.group} and  \ref{table:Binaries} but they
are omitted from the Mass-Temperature relation determination.   

\subsection{Group IV: $\chi$$^2$ Hya}  In this system, the primary is significantly 
evolved beyond the main sequence (Clausen and Nordstrom 1978).  For this reason, it is 
not used in the main sequence  Mass-Temperature relation determination.

\section{Hot Stars} \label{hot}
 One of the aims of this project was to explore the upper range of temperature
where {\it IUE} spectra provide definitive temperatures.  The hottest stars
in the sample (V478 Cyg, AH Cep, and V578 Mon) did not produce well determined
temperature (Section~\ref{systems}).  U Oph also was unsatisfactory because of the
overexposed spectrum 

Ribas et al. (2000) have determined the temperatures from photometry,
compared with Kurucz atmospheres. In order to check the upper
temperature limit for which the {\it IUE} spectra are accurate, the
temperature determinations of the two hottest stars in the {\it IUE}
DEB sample
are compared with the Ribas et al. temperatures in Table~\ref{hotst}.
The temperatures of the two hottest stars agree within the errors. 

\begin{deluxetable}{rlll}
\tablenum{5}
  \tablecaption{Temperatures of Early B Stars \label{hotst}}
\tablewidth{900pt}
\tablehead{
\colhead{      } & \colhead{Spect}   & \colhead{T} & \colhead{T} \\ 
\colhead{   } & \colhead{Type}  & \colhead{Table 4} & \colhead{Ribas et al.}   \\
\colhead{      } & \colhead{}   & \colhead{} & \colhead{(2000)} \\ 
\colhead{      } & \colhead{}   & \colhead{(K)} & \colhead{(K)} \\ 
} 
\startdata
QX Car & B2 V  & 23428 $\pm$ 1000 & 24831 $\pm$ 520  \\
CV Vel & B2.5 V  & 18045 $\pm$ 500  & 17947 $\pm$ 500  \\
\enddata
\end{deluxetable}

\section{Discussion} \label{sec:Our M-T Relation}

\subsection{Binary Parameter Summary}

Parameters for the DEB systems are summarized in  Table~\ref{table:Binaries}.
Each system has two rows, one for the primary and one for the secondary.  The first
entry is the name of the system (line 1) and the HD or BD number (line 2). The next 
columns are  the spectra type, the mass
and its error in $M_{\odot}$,  the temperature and its
error in K, the radius and its error in  $R_{\odot}$, the E(B-V) and its error (where known) in mag, 
the temperature and error from this study (Table~\ref{t.group}), and the log luminosity
in solar units (Section~\ref{l.t}).  
The mass, spectral type, 
radius, temperature, and E(B-V) are from TAG.  E(B-V) for WX Cep is from Graczyk et al (2019).

\begin{deluxetable*}{lcccccccccccc}[p!]
\tablenum{6}
\tablecaption{Parameters for the DEB Sample \label{table:Binaries}}
\tablewidth{500pt}
\tabletypesize{\scriptsize}
\tablehead{
\colhead{System}  & \colhead{SpecType} & \colhead{Mass} & \colhead{$\pm$} & \colhead{T} & \colhead{$\pm$} & \colhead{Radius} &  \colhead{$\pm$} & \colhead{E(B-V)} & \colhead{$\pm$} &  \colhead{T}  & \colhead{$\pm$} &  \colhead{Log Lum}  \\
\colhead{}  & \colhead{} & \colhead{} & \colhead{} & \colhead{TAG} & \colhead{TAG} & \colhead{} & \colhead{} &\colhead{} &\colhead{}  &\colhead{IUE}  &\colhead{IUE} & \colhead{}  \\
\colhead{}  & \colhead{} & \colhead{($M_{\odot}$)} & \colhead{($M_{\odot}$)} & \colhead{(K)} & \colhead{(K)} & \colhead{($R_{\odot}$)} & \colhead{($R_{\odot}$)} &\colhead{(mag)} &\colhead{(mag)}  &\colhead{(K)}  &\colhead{(K)} & \colhead{($L_{\odot}$)}  \\
} 
\startdata
      {\bf   V478 Cyg}    & O9.5V &  16.62  & 0.33 & 30479 &  1000 &  7.426 &  0.072 &  0.85 &  \nodata &  \nodata &  \nodata & \nodata \\    
        HD 193611 &    O9.5V &  16.27 &  0.33 &  30549 &  1000  & 7.426 &  0.072 & \nodata  & \nodata    &\nodata &\nodata &\nodata \\   
    {\bf        AH Cep}  &    B0.5Vn &  15.26 &  0.35 &  29900 &  1000  & 6.346 &  0.071 &  0.58 &  0.08  &   \nodata   &   \nodata  &  \nodata   \\  
        HD 216014 &   B0.5Vn &  13.44 &  0.25 &  28600 &  1000 &  5.836 &  0.085 &     \nodata &    \nodata  &  \nodata  &  \nodata  &  \nodata  \\  
     {\bf     V578 Mon} &    B1V &  14.50 &  0.12 &  30000 &  740 &  5.149 &  0.091 &  0.455 &  0.029 &   \nodata  &    \nodata   &  \nodata  \\
        HD 259135  &   B2V &  10.262 &  0.084  & 26400 &  600 &  4.21 &  0.10 &     \nodata &     \nodata &  \nodata  &  \nodata  &   \nodata \\  
     {\bf     QX Car} &     B2V &  9.25 &  0.12  & 23800 &  500 &  4.291 &  0.091 &  0.049 &   \nodata  &  23428    &  1000   &  3.6701  \\ 
        HD 86118 &   B2V &  8.46 &  0.12 &  22600 &  500 &  4.053 &  0.091 &     \nodata &     \nodata  &   \nodata &  \nodata  &  \nodata  \\  
     {\bf     V539 Ara} &    B3V  & 6.24 &  0.066  & 18100 &  500 &  4.516 &  0.084 &  0.071 &      \nodata &  19248    &  500   &  3.3828  \\ 
        HD 161783 &    B4V &  5.314 &  0.06 &  17100 &  500  & 3.428 &  0.083 &    \nodata  &  \nodata    &  \nodata  &   \nodata &  \nodata  \\  
     {\bf     CV Vel}  & B2.5V &  6.086 &  0.044 &  18100 &  500 &  4.089 &  0.036 &  0.035 &     \nodata & 18045    &   500  &  3.1535  \\ 
        HD 77464 &    B2.5V &  5.982 &  0.035 &  17900 &  500 &  3.950  & 0.036 &     \nodata &     \nodata &  \nodata & \nodata &  \nodata  \\ 
     {\bf     U Oph} &   B5V &  5.273 &  0.091 &  16440  & 250  & 3.484 &  0.021 &  0.226 &  0.007 &    \nodata  &   \nodata      &   \nodata   \\ 
        HD 156247 &   B6V &  4.739 &  0.072 &  15590 &  250 &  3.110 &  0.034 &   \nodata   &     \nodata &   \nodata & \nodata & \nodata  \\  
     {\bf     V760 Sco} &    B4V &  4.969  & 0.09 &  16900 &  500 &  3.015 &  0.066 &  0.33 &    \nodata  &  17092  &  500    &  2.8112  \\
        HD 147683 &   B4V &  4.609 &  0.073 &  16300 &  500 &  2.641 &  0.066 &     \nodata &   \nodata   &  \nodata  \\ 
      {\bf    GG Lup} &    B7V &  4.106 &  0.044 &  14750 &  450 &  2.380 &  0.025 &  0.027 &     \nodata &   15793  &  250   &  2.3995  \\
        HD 135876 &    B9V &  2.504 &  0.023 &  11000 &  600 &  1.726 &  0.019 &    \nodata  &     \nodata &   \nodata  &  \nodata  &   \nodata  \\ 
        {\bf  $\zeta$ Phe} &    B6V & 3.921  & 0.045 & 14400  & 800  & 2.852  & 0.015  & 0  &     \nodata &  14483   &   250 &   2.5539  \\ 
     HD 6882 &    B8V & 2.545  & 0.026  & 12000   & 600  & 1.854  & 0.011  &     \nodata &     \nodata &  \nodata   &   \nodata &   \nodata  \\ 
     {\bf     $\chi^ 2 $ Hya}  &  B8V & 3.605 & 0.078 & 11750 & 190 & 4.390 & 0.039 & 0.016 &    \nodata &  12160  & 250   &  2.4619  \\
        HD 96314 &  B8V & 2.632 & 0.049 & 11100 & 230 & 2.159 & 0.030 &    \nodata &    \nodata  &   \nodata &  \nodata  &   \nodata \\
      {\bf    IQ Per }  & B8V & 3.504 & 0.054 & 12300 & 230 & 2.445 & 0.024 & 0.14 & 0.01  &  13241   &  250   &  2.1479  \\
        HD 24909  & A6V & 1.730 & 0.025 & 7700  & 140 & 1.499 &  0.016 &     \nodata &     \nodata &   \nodata \\  
      {\bf    PV Cas }  & B9.5V & 2.816 & 0.05 & 10200 & 250 & 2.301 & 0.020 & 0.217 &    \nodata &   10632   &   250   &  1.7687   \\
        HD 240208  & B9.5V & 2.757 & 0.054 & 10190 & 250 & 2.257 & 0.019 &    \nodata &    \nodata  &  \nodata  &  \nodata  &   \nodata \\
     {\bf     V451 Oph} &   B9V &  2.769 &  0.062 &  10800 &  800 &  2.642 &  0.031 &  0.158 &     \nodata  &  11980  &   250  & 1.9555 \\ 
        HD 170470 &    A0V &  2.351 &  0.052 &  9800 &  500 &  2.029 &  0.028 &     \nodata &     \nodata &  \nodata  &  \nodata  &  \nodata  \\ 
      {\bf    WX Cep} &   A5V &  2.533 &  0.05 &  8150 &   250 &  3.996 &  0.03 &  0.19 &  0.03  &     \nodata    &     \nodata     & \nodata  \\ 
        HD 213631 &   A2V &  2.324 &  0.045 &  8900 &  250  & 2.712 &  0.023 &     \nodata &     \nodata &  9087   &  125  &   \nodata  \\ 
     {\bf     TZ Men } &   A0V &  2.482 &  0.025 &  10400 &  500 &  2.017 &  0.02 &  0 &     \nodata  &  10418  & 100   & 1.6741    \\ 
        HD 39780 &   A8V &  1.500 &  0.010 &  7200 &  300  &  1.433 &  0.014 &     \nodata &    \nodata   &   \nodata &  \nodata  &  \nodata  \\ 
     {\bf     V1031 Ori} &   A6V &  2.468 &  0.018 &  7850 &  500 &  4.323 &  0.034 &  0.034 &     \nodata  &     \nodata &   \nodata  &    \nodata  \\  
        HD 38735 &    A3V &  2.281 &  0.016 &  8400 &  500 &  2.978 &  0.064 &     \nodata &     \nodata & 8906   & 100   & \nodata  \\ 
     {\bf     $\beta$ Aur} &    A1m &  2.375 &  0.027 &  9350 &  200  & 2.765 &  0.018 &  0 &     \nodata &  9177    &  125  &     \nodata \\
        HD 40183 &   A1m &  2.304 &  0.030 &  9200 &  200 &  2.571 &  0.018 &     \nodata &     \nodata  &   \nodata &  \nodata  &  \nodata  \\ 
      {\bf    YZ Cas} &    A1m &  2.317 &  0.020 &  10200 &  300 &  2.539 &  0.026 &  0.07 &    \nodata  &  8982  &  125    &  1.7423  \\ 
        HD 4161 &     F2V &  1.352 &  0.009 &  7200 &  300 &  1.350 &  0.014 &     \nodata &    \nodata   &  \nodata  &  \nodata  &  \nodata  \\
     {\bf     V624 Her}  &  A3m &  2.277 &  0.014 &  8150 &  150  & 3.031 &  0.051 &  0.05 &  0.01 &   8160   &  125  &  1.5924  \\
        HD 161321 &   A7m: &  1.876  & 0.013  & 7950 &  150 &  2.210 &  0.034 &     \nodata &     \nodata &   \nodata &   \nodata &   \nodata \\ 
     {\bf     GZ CMa}  &  A3m &  2.199 &  0.017 &  8800 &  350 &  2.494 &  0.031 &  0.07 &  0.03 &  8566  &  125    &   1.4951 \\
        HD 56429 &    A4V: &  2.006 &  0.012 &   8500 &  350 &  2.132 &  0.037 &     \nodata &     \nodata  &  \nodata  &   \nodata &  \nodata  \\ 
      {\bf    V1647 Sgr} &   A1V &  2.184 &  0.037 &  9600 &  300 &  1.832 &  0.018 &  0.04 &     \nodata  &  9361   &  125     & 1.4346  \\
        HD 163708 &    A1V &  1.967 &  0.033 &  9100 &  300 &   1.667 &  0.017 &     \nodata &     \nodata &  \nodata  &  \nodata  &   \nodata \\ 
     {\bf     EE Peg} &    A3m &  2.151 &  0.024 &  8700 &  200 &  2.090 &  0.025 &  0 &     \nodata &  8687   &  125   &   1.2871 \\   
        HD 206155 &  F5V & 1.332 & 0.011 & 6450 & 300 & 1.312 & 0.013 &   \nodata &   \nodata  & \nodata  & \nodata  & \nodata  \\
     {\bf     VV Pyx }  & A1V & 2.097 & 0.022 & 9500 & 200 & 2.168 & 0.02 & 0.022 &    \nodata &  9751   &  125   &  1.7809 \\
        HD 71581  & A1V & 2.095 & 0.019 & 9500 & 200 & 2.168 & 0.02 &    \nodata &    \nodata  &   \nodata &  \nodata  &  \nodata  \\
       {\bf    AY Cam}  & A0V & 1.905 & 0.04 & 7250 & 100 & 2.772 & 0.02 & 0.06 &    \nodata &    \nodata   &   \nodata    & \nodata  \\
        BD+77 328 &  F0V & 1.707 & 0.036 & 7395 & 100 & 2.026 & 0.017 &  \nodata &  \nodata & 7590  &  150 &  \nodata \\
\enddata
\end{deluxetable*}


\subsection{Comparison between T$_{IUE}$ and  T$_{TAG}$}
\begin{figure}
\plotone{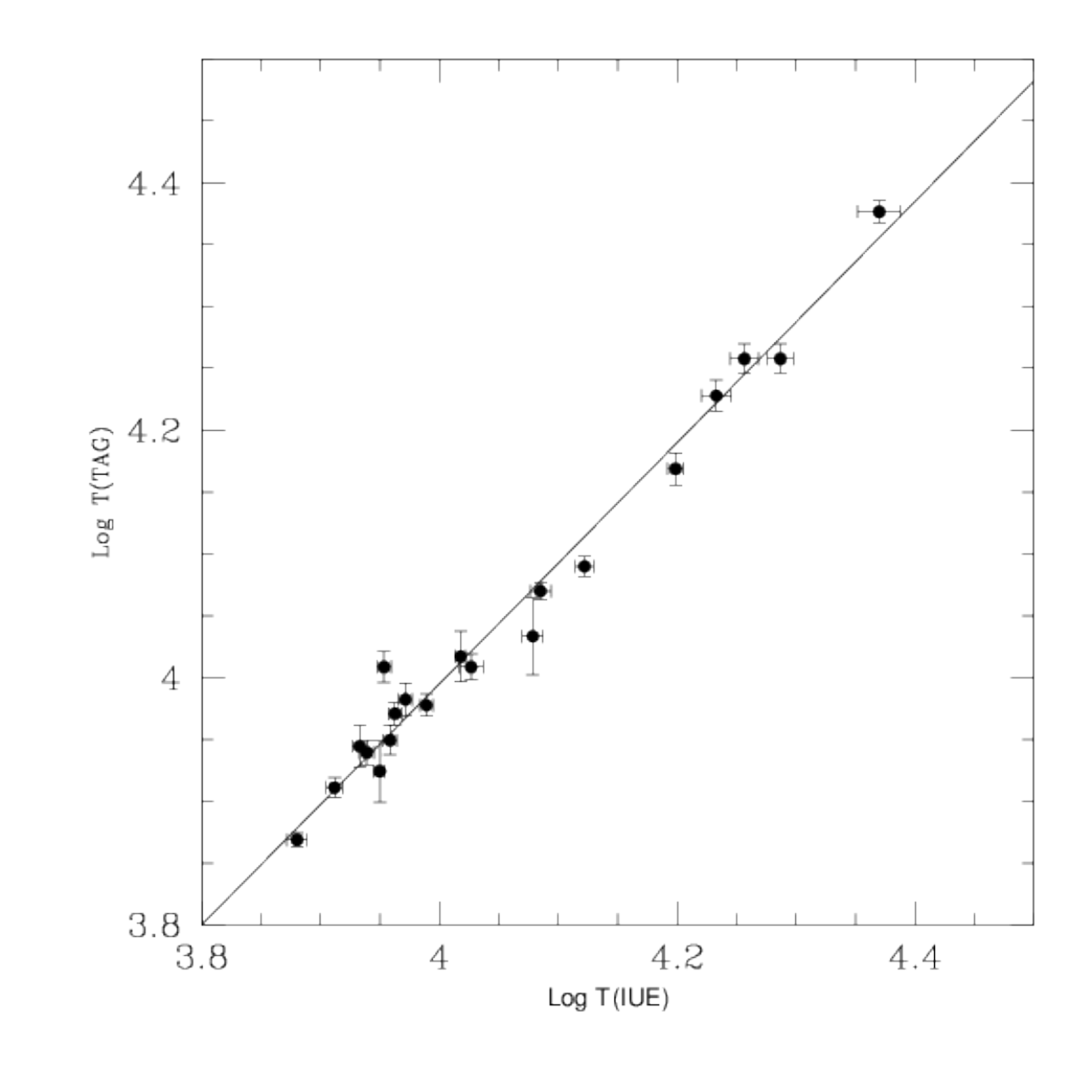}
  \caption{The temperatures from the {\it IUE} spectral fit 
Log T(IUE) versus the temperatures from TAG Log T(TAG).  Both temperatures are in K.
\label{tt.comp}}
\end{figure}

The comparison between the log of the IUE temperatures derived here from
 spectral fits  and the log of the 
temperatures from TAG is shown in 
Fig.~\ref{tt.comp}.  In general they are similar within the errors, with an 
insignificant offset.  The line shows the fit 
from the least squares bisector (unweighted), as recommended by Isobe, et al. (1990):
 
log T(TAG) = 0.104 $\pm$ 0.114 $ + $ (0.973 $ + $ 0.020) $\times$ log T(IUE)  (Eqn 1)

\subsection{The M-R Relation: Evolution}
\begin{figure}
\plotone{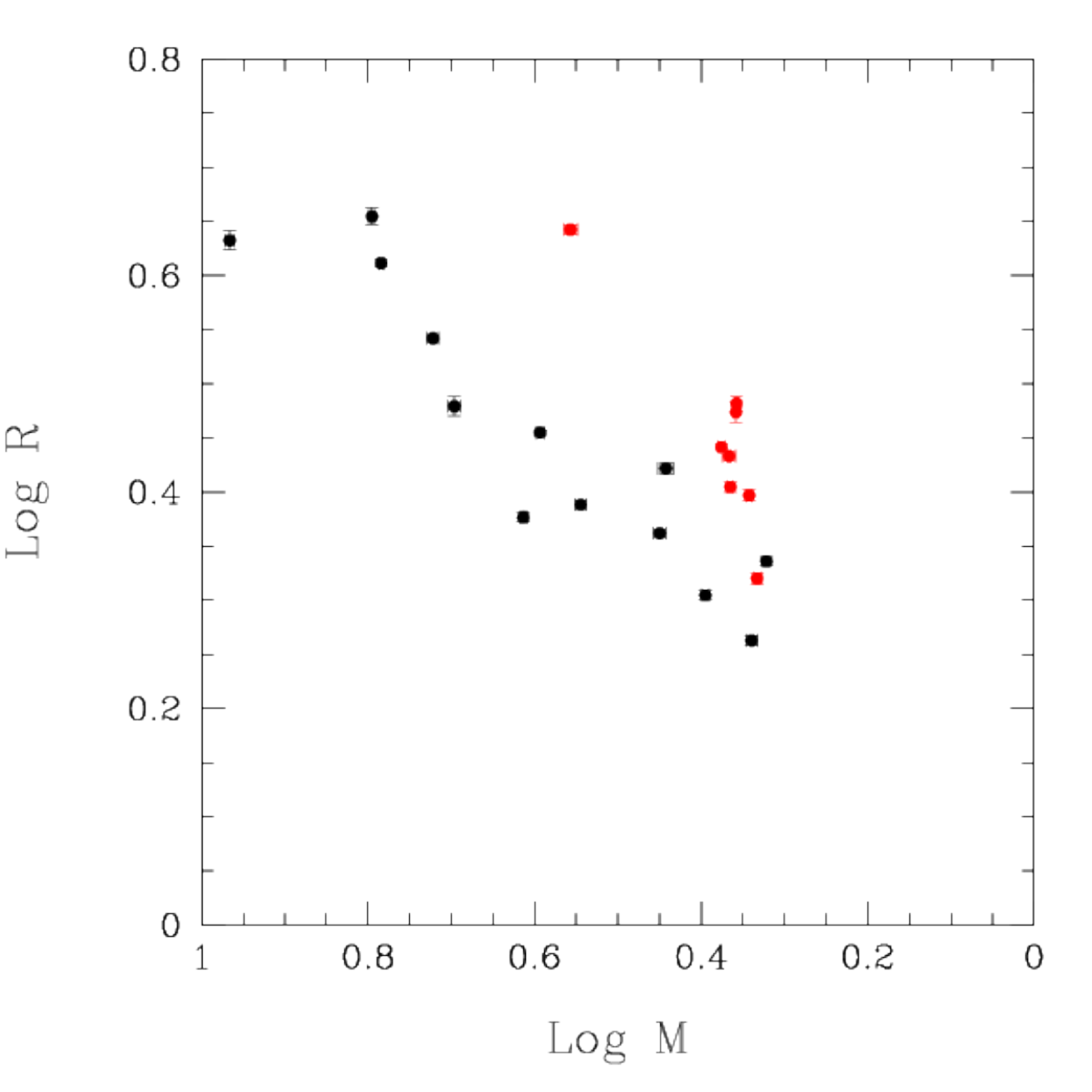}
\caption{Log Radius vs Log Mass relation from the TAG sample.  
Main sequence stars are in black; other stars are in red. The 
group of other stars includes $\chi^2$ Hya, Group III stars, and Am stars (see text).
 Masses are in $M_{\odot}$; radii are in  $R_{\odot}$
\label{m.r}}
\end{figure}

The Log Mass--Log Radius relation from TAG data is shown in Fig.~\ref{m.r}.  
Stars with radii larger than the main 
sequence minimum indicates evolution beyond the Zero Age Main Sequence, 
which is a major source of scatter in the Mass-Temperature relation. 
The stars marked in red in  Fig.~\ref{m.r} are from several groups identified in this study. 
They include $\chi^2$ Hya, which is known to be evolved.  Group III stars 
are also evolved.  In this case 
it is the temperature of the secondary which we determine.    The 
primary is clearly evolved in Group III systems,
 but the secondary may also be.  In fact, for all three stars 
(WX Cep, V1031 Ori and AY Cam), the
radius is larger than the radii for the main sequence.  The 5 stars with an Am
 (metallic) primary ($\beta$ Aur, YZ Cas,
V624 Her, GZ CMa, and EE Peg) are discussed in the next section. All these
stars shown in red  sit above the
main sequence relation in  Fig.~\ref{m.r}, and account for most of the spread in the Log M--Log R 
relation.    This is discussed further in the next section.  

\subsection{Metallicity}
Five stars in the DEB sample have a metallic Am primary ($\beta$ Aur, YZ Cas,
V624 Her, GZ CMa, and EE Peg). In the previous section and Fig.~\ref{m.r}, it was 
shown that they have larger radii (for a given mass) than normal A stars.  In this
case, they are not necessarily evolved, simply showing the effects of a metal-rich  
atmosphere.  In the mass-temperature relation  
in the next section,  they typically have cooler temperatures than normal A 
stars coupled with their larger radii.  Again, this is explained by a metal-rich 
atmosphere.  In addition, the BOSZ models used here assume  solar abundances, 
and hence are not appropriate for Am stars. For this reason these stars are omitted
from the determination of the Mass-Temperature relation.   

\subsection{The M-T Relation} 
\begin{figure}
\plotone{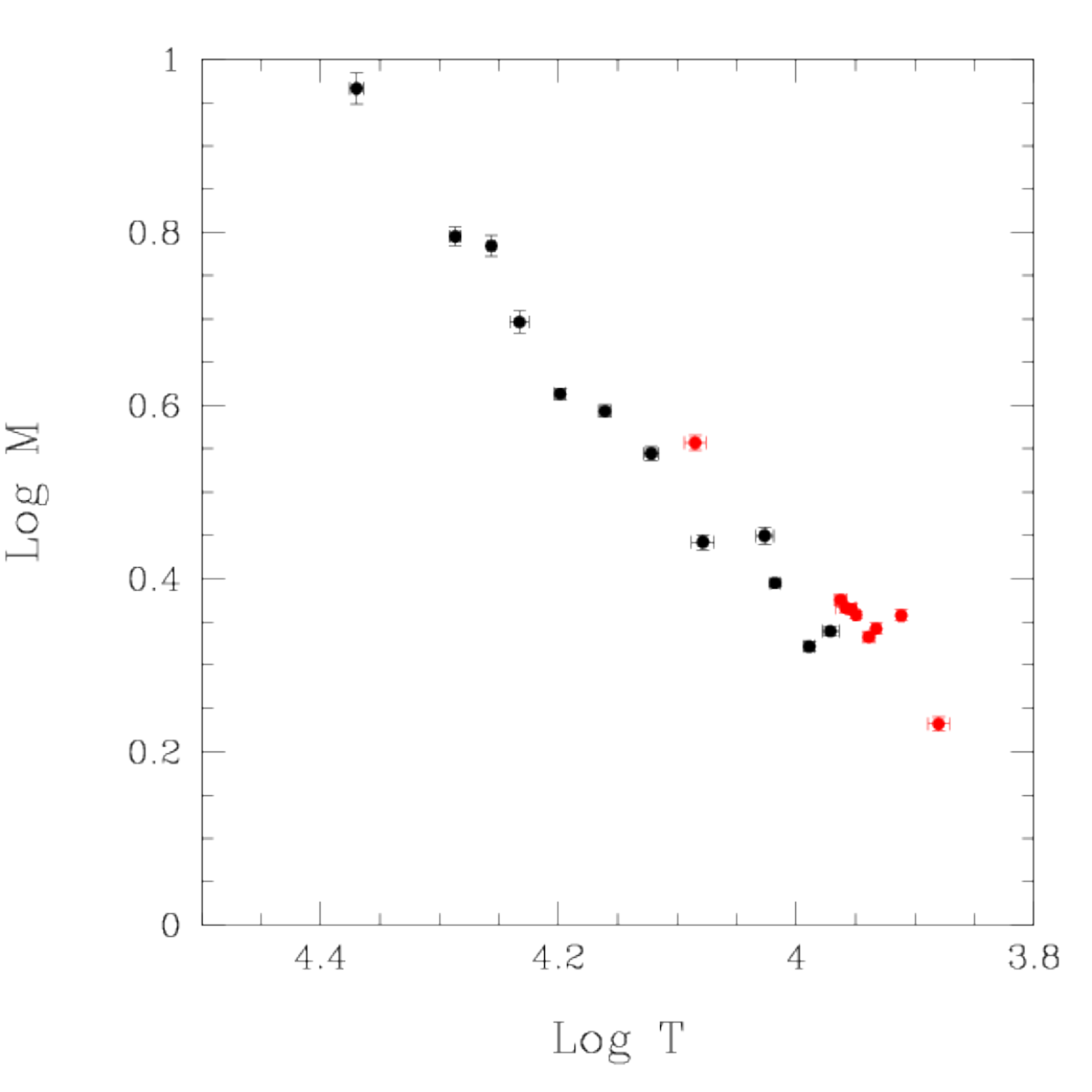}
  \caption{The Mass -Temperature relation. Main sequence stars: black; others: red. Others include
$\chi^2$ Hya, Group III stars, and Am stars. Masses are in $M_{\odot}$; temperatures are in K.
     \label{m.t}}
\end{figure}

The relation between mass and temperature is shown in 
Fig.~\ref{m.t}.  (The errors in the temperatures are from the visual inspection.) The systems
plotted in black are used to determine the relation (omitting stars showing evolution and 
chemically peculiar stars plotted in red).

For this limited temperature (23000 to 8000K) and mass range, we have assumed a linear relation between
log M and log T.
The relation from bisector least squares is: 

log M/$M_{\odot}$ =  -5.90 $\pm$ 0.27 + (1.56 $\pm$  0.07) x log T   (Eqn 2)

 The RMS of the residuals from the fit
  of the log  M/$M_{\odot}$ is 0.037. 

   This mass--temperature relation is directly determined. However, there are several
    factors which contribute to the width of the relation.  Different rotation
    velocities and different abundances will result in a spread in temperatures
    between stars.  In addition during their main sequence lifetimes, stars cool,
    resulting in a range of temperatures.  As an example, the evolutionary
    tracks from Ekstr\"om, et al. (2012) span approximately 0.1 dex in Log T during
    the main sequence.  This corresponds to an uncertainty of approximately 0.08 dex
    in log M or about 18\%.   The scatter about the mean relation ($\pm$ 0.04 dex in log M)
    due to evolution within the main sequence is about the same as the RMS around
    equation 2.  That is, the age spread accounts for at least a large part of the
    RMS scatter.

    In certain cases, this age-related uncertainty is lower.  For instance,  a secondary which is the
    companion of a more massive star must be comparatively young.  The hot companion of
    a Cepheid is  an example which must be close to the zero age main sequence (ZAMS).
    That is, it will  be on the hotter edge of the M-T relation with a smaller
  uncertainty than a random main sequence star.

\subsection{Comparisons Against Other M-T Relations}

\begin{figure}
\plotone{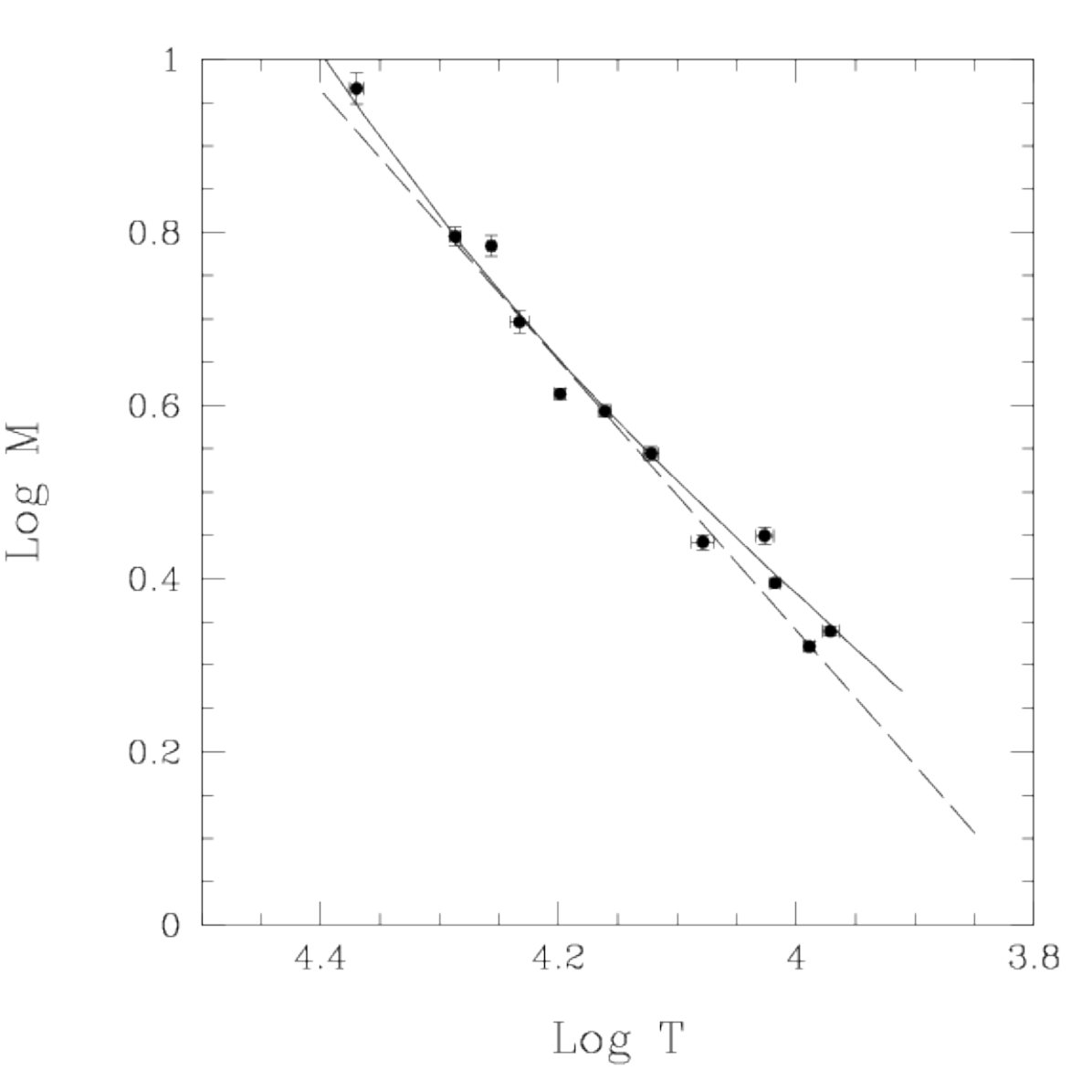}
  \caption{The Mass -temperature relation. Main sequence stars: black points; dashed line: fit to
points; solid line: Harmanec relation.
     \label{m.t.comp}}
\end{figure}

The linear fit to the main sequence points (black points in Fig.~\ref{m.t})
is shown in Fig.~\ref{m.t.comp} in the dashed line.
Harmanec (1988) used the sample of DEBs available 
at that time to derive a Mass-Temperature relation over
the full length of the main sequence. The
solid line shows the relation from Harmanec (1988).  The Harmanec relation is curved since it describes 
a much larger temperature range.  However the agreement is essentially good in this temperature range. 

\subsection{The Luminosity--Temperature Relation} \label{l.t}

\begin{figure}
\plotone{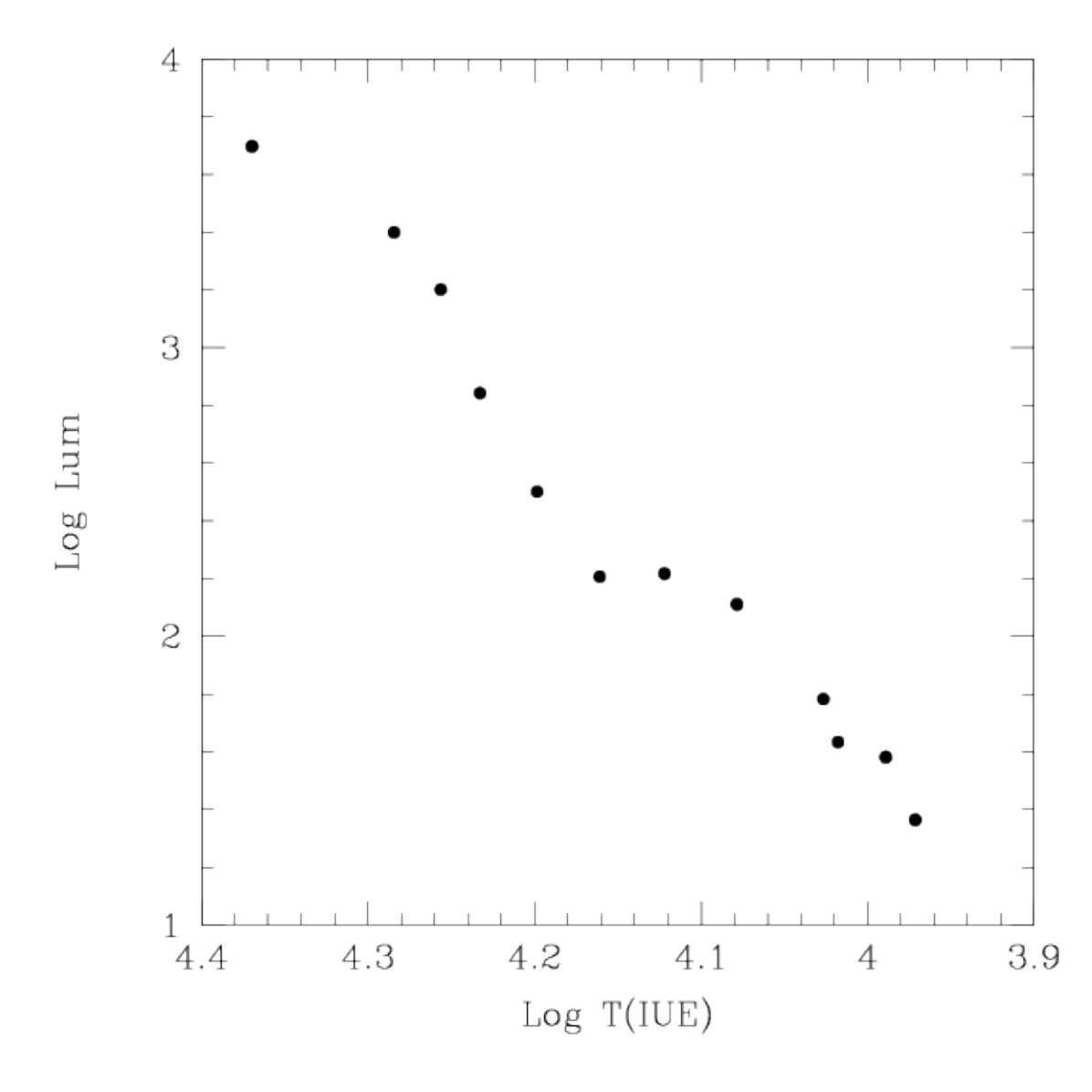}
  \caption{The Log Temperature--log Luminosity relation for main sequence stars showing
no evolution or chemical peculiarities (see text). 
Luminosity is in solar units; the temperature is
determined from the IUE spectral fits.
     \label{t.lum}}
\end{figure}

For completeness we provide the Log Temperature --Log Luminosity relation
in Fig. ~\ref{t.lum}.
Luminosities are derived from TAG values scaled by $(T_{IUE}/T_{TAG})^4$.  
The stars shown 
omit systems showing evolution ($\chi^2$ Hya, WX Cep, V1031, and AY Cam), as well 
as Am stars. 


\section{Conclusions} \label{sec:Conclusion}
This study combines masses for DEBs with temperatures determined from ultraviolet IUE spectra 
and BOSZ model atmospheres.  For this 
important group on which the calibration of masses on the main sequence rests, a 
direct Mass-Temperature relation is determined for 
B and early A stars. The mass-radius plot (Fig~\ref{m.r})
shows that some of the sample is evolved beyond the main sequence. These stars
as well as chemically peculiar Am stars have been removed from the calibration sample. 
The  Mass-Temperature relation 
 is probably still somewhat broadened by variation in evolution and abundance,  however  
 Fig.~\ref{m.t} demonstrates that the scatter is relatively small
(RMS of the log  M/$M_{\odot}$ is 0.037). 
 The comparison between the IUE 
temperatures and those in TAG from spectral types is good.  Furthermore, the  Mass-Temperature 
relation is very similar to that
derived by Harmanec (1988).    


\section{Acknowledgments}
The authors would like to thank  Dr. Hans Moritz G\"unther for his insightful comments and 
assistance regarding our temperature fitting. We thank the referee for comments that
improved the content of the paper as well as the presentation. 
It is a pleasure to thank Dr. Richard Gray for
comments on chemically peculiar stars. 
 M.G.F would like to thank Dr. Matt Ashby and 
Dr. Jonathan McDowell for their invaluable comments and guidance during the SAO REU program. 
The SAO REU program is funded in part by the National Science Foundation REU and Department 
of Defense ASSURE programs under NSF Grant no. AST-2050813, and by the Smithsonian 
Institution.  Support was provided to N.R.E. by HST Grant GO-15861.001-A and
 by the Chandra X-ray Center NASA Contract NAS8-03060.
This research is based on observations made with the {\it IUE} satellite, 
obtained from the MAST data archive at the Space Telescope Science Institute, 
which is operated by the Association of Universities for Research in Astronomy, Inc., 
under NASA contract NAS 5–26555.
This research has made use of the SIMBAD database, operated at CDS, Strasbourg, France.

\section{Appendix A: Early B Stars}

In order to confirm the temperature sensitivity of the IUE spectra of the hotter stars in 
the DEB sample we examined stars from the {\it IUE} Spectral Atlas (Wu, et al. 1983).
The {\it IUE} Atlas project observed 
samples of the full range of the Morgan-Keenan (MK) spectral system in low dispersion in both the 
short and long wavelength regions, primarily using stars defining the MK classes. Table 7
lists the spectra selected for this test for spectral classes B1 V, B3 V, and B5 V.  
Columns provide the spectral type, the HD number, V, E(B-V), the {\it IUE} spectrum number, 
the exposure level, alternate designations,
 MK for stars which define MK classes, and the temperature, and the temperature 
uncertainty from the SD comparisons, and from
visual comparisons.  The maximum exposure is in instrumental Data Numbers (DN),
where an optimal exposure is near 200 DN. 
 The spectra were fitted to BOSZ models (1T fit). 
The resulting temperatures are listed in Table 7.   The final three
columns list temperatures for these classes for 3 widely used 
calibrations of temperatures for MK classes: Drilling and Landolt in Astrophysical 
Quantities (2000), Harmanec (1988) and Pecaut and Mamajek (2013). 
 The agreement is reasonable, 
 given that MK classes have some width in temperature.
  Lyubimkov et al. (2002)
derive T=22500 $\pm$ 600 K for HD31726, which is close to
the {\it IUE} result

\begin{deluxetable*}{lrlrlllllrrrrr}
\tablenum{7}
\tablecaption{\textit{IUE} Observations of MK Standard Stars \label{stds}}
\tablewidth{900pt}
\tablehead{
\colhead{Spectral} & \colhead{HD} & \colhead{V} & 
\colhead{E(B-V)} & \colhead{Spectrum} & 
\colhead{Exp Max} & \colhead{ID} & \colhead{} &
 \colhead{T} & 
\colhead{$\pm$} & \colhead{$\pm$} & \colhead{AQ} & \colhead{Harm}& \colhead{PM}   \\ 
\colhead{Type} & \colhead{} & \colhead{} & \colhead{} & 
\colhead{} & \colhead{} & \colhead{} & \colhead{} 
 & \colhead{} &
\colhead{SD} &  \colhead{Vis}  &  \colhead{}&  \colhead{} &  \colhead{}  \\
\colhead{} & \colhead{} & \colhead{(mag)} & \colhead{(mag)} & 
\colhead{} & \colhead{(DN)} & \colhead{} & \colhead{} 
& \colhead{(K)} &
\colhead{(K)} &  \colhead{(K)}  &  \colhead{(K)}  &  \colhead{(K)}  &  \colhead{(K)} \\
} 
\startdata
B1 V & 31726 & 6.15  & 0.05 &   SWP8165 &  230 &    &    & 22471.0 &  908 & 500 &  24540  & 26180 & 26000 \\
B3 V & 190993 & 5.06 &  0.02 &   SWP9961 &  200  &  17 Vul &  MK &  18025.1 & 401 & 500 & 19000 & 19050  & 17000  \\
B5 V &  34759 &  5.22 &  0.02 &   SWP15537 & 190 &   $\rho$ Aur & MK &  16373.3 &  378  & 375 & 15200 & 15490 & 15700 \\
\enddata

 AQ: Drilling,  and Landolt  (2000)

Harm:  Harmanec (1988)

PM:  Pecaut and Mamajek (2013)

\end{deluxetable*}







\begin{thebibliography}{}



\bibitem[\protect\citeauthoryear{andersen}{1983}]{ander83}Andersen, J.\ 1983,  A\&A, 118, 255


\bibitem[\protect\citeauthoryear{andersen etal}{1983}]{and83}Andersen, J., Clausen, J.~V., Nordstroem, B., et al.\ 1983, A\&A, 121, 271

\bibitem[\protect\citeauthoryear{andersen etal}{1993}]{and93}Andersen, J., Clausen, J.~V., \& Gimenez, A.\ 1993,  A\&A, 277, 439

  
\bibitem[\protect\citeauthoryear{andersen etal}{1985}]{and85}
  Andersen, J., Clausen, J.~V., Nordstrom, B., et al.\ 1985, A\&A, 151, 329

\bibitem[\protect\citeauthoryear{andersen etal}{1984}]{and84}
Andersen, J., Clausen, J.~V., \& Nordstrom, B.\ 1984, A\&A, 134, 147

\bibitem[\protect\citeauthoryear{andersen etal}{1987}]{and87}
  Andersen, J., Clausen, J.~V., \& Nordstrom, B.\ 1987,  A\&A, 175, 60

\bibitem[\protect\citeauthoryear{andersen etal}{1990}]{and90}
  Andersen, J., Clausen, J.~V., \& Nordstrom, B.\ 1990,  A\&A, 228, 365

\bibitem[\protect\citeauthoryear{andersen gimenez}{1985}]{andgim85}
Andersen, J. \& Gimenez, A.\ 1985, A\&A, 145, 206


\bibitem[\protect\citeauthoryear{andersen}{1991}]{and91}
Andersen, J. 1991, A\&ARv, 3, 91.


\bibitem[\protect\citeauthoryear{barembaum etzel}{1995}]{bar85}
 Barembaum, M.~J. \& Etzel, P.~B.\ 1995, AJ, 109, 2680

\bibitem[\protect\citeauthoryear{bohlin bianchi}{2017}]{bol17}
 Bohlin R. C. and Bianchi, L 2017, AJ, 155, 162

\bibitem[\protect\citeauthoryear{bohlin etal}{2017}]{bmf17}
  Bohlin, R. C., Meszaros, S,, Fleming, S. W., Gordon, K. D., Koekemoer,
  A. M. and Kovacs, J. 2017, AJ, 153, 23

\bibitem[\protect\citeauthoryear{cardelli etal}{1989}]{ccm89}
 Cardelli, Clayton, and Mathis 1989, ApJ  345,  245

\bibitem[\protect\citeauthoryear{clausen}{1996}]{cla96}
 Clausen, J.~V.\ 1996, A\&A, 308, 151


\bibitem[\protect\citeauthoryear{clausen etal}{1986}]{cla86}
Clausen, J.~V., Gimenez, A., \& Scarfe, C.\ 1986, A\&A, 167, 287

\bibitem[\protect\citeauthoryear{clausen nordstrom}{1978}]{cla78}
Clausen, J. V. and Nordstrom, B. 1978, A\&A, 67, 15

\bibitem[\protect\citeauthoryear{drilling landolt}{2000}]{dri00}
Drilling, J. S., and Landolt, A. U. in Cox, A. N. 2000, Astrophysical Quantities, Springer: London, 388

\bibitem[\protect\citeauthoryear{ekstrom etal}{2012}]{eks12}
Ekstr\"om, S., Georgy, C. and Eggenberger, P. et al. 2012, A\&A, 537, A146

\bibitem[\protect\citeauthoryear{garcia etal}{2013}]{gar13}
Garcia, E. V., Stassun, K. G., and Torres, G. 2013, ApJ, 769, 114

\bibitem[\protect\citeauthoryear{Graczyk etal}{20}]{gra19}
 Graczyk D., Pietrzynski G., Gieren W. et al. 2019, ApJ, 872, 85

\bibitem[\protect\citeauthoryear{Harmanec}{88}]{har88}
Harmanec, P. 1988, Bull. Astron. Inst. Czechosl, 39, 329

\bibitem[\protect\citeauthoryear{harris sonneborn}{1987}]{hs87}
Harris, A. W. and Sonneborn, G. 1987, in Exploring the Universe with the IUE 
Satellite, ed. Y. Kondo, et al. (Reidel Publishing Company, 
Dortrecht, Holland), 729 

\bibitem[\protect\citeauthoryear{hensberge etal}{2000}]{hen00}
Hensberge, H., Pavlovski, K., and Verschueren, W. 2000 A\&A, 358, 553

\bibitem[\protect\citeauthoryear{holmgren etal.}{1990}]{hol90}
Holmgren, D.~E., Hill, G., and Fisher, W.\ 1990,  A\&A, 236, 409


\bibitem[\protect\citeauthoryear{isobe etal}{1990}]{iso90}
Isobe, T., Feigelson, E. D., Akritas, M. G., and Babu, G. J. 1990, ApJ, 364, 104

\bibitem[\protect\citeauthoryear{lacy popper}{1984}]{lac84}
Lacy, C.~H. and Popper, D.~M.\ 1984, ApJ, 281, 268

\bibitem[\protect\citeauthoryear{lyubimkov etal}{2002}]{lyu02}
Lyubimkov, L. S., Rachkovskaya, T. M., Rostopchin, S. I., and
Lambert, D. L. 2002, MNRAS, 333, 9

\bibitem[\protect\citeauthoryear{nichols linsky}{1996}]{nic96}
Nichols, J. S. and Linsky, J. L. 1996, AJ, 111, 517

\bibitem[\protect\citeauthoryear{pavlovski etal}{2014}]{pav14}
Pavlovski, K., Southworth, J., Kolbas, V., et al.\ 2014, MNRAS, 438, 590


\bibitem[\protect\citeauthoryear{pecaut mamajek}{2013}]{pec13}
Pecaut, M. J. and Mamajek, E. E. 2013, ApJS, 208, 9

\bibitem[\protect\citeauthoryear{popper}{1987}]{pop87}
  Popper, D.~M.\ 1987, AJ, 93, 672

\bibitem[\protect\citeauthoryear{popper dumont}{1977}]{pop77}
Popper, D. M. and Dumont, P. J. 1977, AJ, 82, 216

\bibitem[\protect\citeauthoryear{popper etzel}{1981}]{pop81}
Popper, D. M. and Etzel, P. B. 1981, AJ, 86,102    

\bibitem[\protect\citeauthoryear{popper}{1984}]{pop84}
Popper, D.~M.\ 1984, AJ, 89, 1057

\bibitem[\protect\citeauthoryear{popper etal}{1985}]{pop85}
Popper, D.~M., Andersen, J., Clausen, J.~V., et al.\ 1985, AJ, 90, 1324

\bibitem[\protect\citeauthoryear{ribas etal}{2000}]{rib00}
Ribas, I., Jordi, C., Torra, J., and Gimenez, A. 2000, MNRAS, 313, 99

\bibitem[\protect\citeauthoryear{southworth etal}{2007}]{sou07}
Southworth, J., Bruntt, H., \& Buzasi, D.~L.\ 2007, A\&A, 467, 1215

\bibitem[\protect\citeauthoryear{torres etal}{2010}]{tor10}
Torres, G., Andersen, J., and Gimenez, A. 2010, A\&ARv, 18, 67  (TAG)

\bibitem[\protect\citeauthoryear{vaz etal}{(2007}]{vaz07}
  Vaz, L.~P.~R., Andersen, J., \& Claret, A.\ 2007, A\&A, 469, 285

\bibitem[\protect\citeauthoryear{williamon etal}{(2004}]{wil04}
Williamon, R.~M., Sowell, J.~R., \& Van Hamme, W.\ 2004, AJ, 128, 1319

\bibitem[\protect\citeauthoryear{Wolf et al.}{(2006}]{wol06}
  Wolf, M., Ku{\v{c}}{\'a}kov{\'a}, H., Kolasa, M., et al.\ 2006, A\&A, 456, 1077. 

  
\bibitem[\protect\citeauthoryear{wu etal}{1983}]{wu83}
Wu, C.-C., Ake, T. B., Boggess, A. et al. 1983, IUE Newsletter \#22    

\bibitem[\protect\citeauthoryear{yakut etal}{2007}]{ya07}
  Yakut, K., Aerts, C., \& Morel, T.\ 2007, A\&A, 467, 647


\end{thebibliography}
\end{document}